\documentclass[aip,jmp,reprint,onecolumn]{revtex4-1}
\usepackage{amsmath}
\usepackage{amssymb}
\usepackage{braket}
\usepackage[usenames,dvipsnames]{color}
\usepackage{enumerate}
\usepackage{graphicx}
\usepackage{hyperref}
\usepackage{listings}
\usepackage{mathtools}
\usepackage{paralist}
\usepackage{rotating}
\usepackage[caption=false]{subfig}
\usepackage[vcentermath]{youngtab}

\allowdisplaybreaks[1]

\newcommand{\SU}[1]{\ensuremath{\text{SU}(#1)}}
\newcommand{\su}[1]{\ensuremath{\mathfrak{su}(#1)}}
\newcommand{\SLC}[1]{\ensuremath{\text{SL}(#1,\mathbb{C})}}
\newcommand{\slc}[1]{\ensuremath{\mathfrak{sl}(#1,\mathbb{C})}}

\newcommand{\gt}[6]{\ensuremath{\begin{pmatrix} \multicolumn{2}{c}{#1} & \multicolumn{2}{c}{#2} & \multicolumn{2}{c}{#3} \\ & \multicolumn{2}{c}{#4} & \multicolumn{2}{c}{#5} & \\ && \multicolumn{2}{c}{#6} && \end{pmatrix}}}

\newcommand{\GAT}{GT}
\newcommand{\GT}{GT}

\newcommand{\GTP}{GT-pattern}

\newcommand{\CGC}{CGC}

\newcommand{\iweight}{i-weight}
\newcommand{\pweight}{p-weight}

\DeclareMathOperator{\Span}{span}

\begin{document}
\title{A numerical algorithm for the explicit calculation of \SU N and \SLC N Clebsch-Gordan coefficients}
\author{Arne Alex}
\email{arne.alex@physik.lmu.de}
\affiliation{Physics Department, Arnold Sommerfeld Center for Theoretical Physics and Center for NanoScience, Ludwig-Maximilians-Universit\"at M\"unchen, D-80333 M\"unchen, Germany}
\author{Matthias Kalus}
\author{Alan Huckleberry}
\affiliation{Fakult\"at f\"ur Mathematik, Ruhr-Universit\"at Bochum, D-44780 Bochum, Germany}
\author{Jan von Delft}
\affiliation{Physics Department, Arnold Sommerfeld Center for Theoretical Physics and Center for NanoScience, Ludwig-Maximilians-Universit\"at M\"unchen, D-80333 M\"unchen, Germany}
\date{31 August 2010}
\begin{abstract}
  We present an algorithm for the explicit numerical calculation of \SU N
  and \SLC N Clebsch-Gordan coefficients, based on the
  \emph{Gelfand-Tsetlin pattern} calculus.
  Our algorithm is well suited for numerical implementation; we include
  a computer code in an appendix. Our exposition presumes
  only familiarity with the representation theory of \SU 2.
\end{abstract}
%
%
\maketitle 
\section{Introduction}
\label{sec:intro}
Clebsch-Gordan coefficients (\CGC{}s) arise when decomposing the
tensor product $\mathbb{V}^{S} \otimes \mathbb{V}^{S'}$ of the
representation spaces of two irreducible representations (\emph{irreps})
$S$ and $S'$ of some group into a direct sum
$\mathbb{V}^{S''_1} \oplus \cdots \oplus \mathbb{V}^{S''_r}$ of irreducible
representation spaces. They describe the corresponding basis transformation
from a tensor product basis $\{\ket{M \otimes M'}\}$ to a basis
$\{\ket{M''}\}$ which explicitly accomplishes this decomposition.

\CGC{}s are familiar to physicists in the
context of angular momentum coupling, in which the direct product of
two irreps of the \SU 2 group is
decomposed into a direct sum of irreps. \SU 3 Clebsch-Gordan
coefficients arise, for example, in the context of quantum chromodynamics,
while \SU N \CGC{}s, for general $N$, appear in the construction of
unifying theories whose symmetries contain the $\SU 3 \times \SU 2 \times U(1)$
standard model as a subgroup\cite{Slansky1981}.
\SU N \CGC{}s are also useful for the numerical treatment of models with
\SU N symmetry, where they arise when exploiting the Wigner-Eckart
theorem to simplify the calculation of matrix elements of the
Hamiltonian. 
Such a situation arises, for example, in the numerical treatment of
\SU N-symmetric quantum impurity models using the numerical renormalization
group \cite{Toth2008}. Such models can be mapped onto \SU N-symmetric,
half-infinite quantum chains, with hopping strengths that decrease
exponentially along the chain. The Hamiltonian is diagonalized numerically
in an iterative fashion, requiring the explicit calculation of matrix elements
of the Hamiltonian of subchains of increasing length. The efficiency of this
process can be increased dramatically by exploiting the Wigner-Eckart theorem,
which requires knowledge of the relevant Clebsch-Gordan coefficients.
(Details of how to implement \SU N symmetries within the context of the
numerical renormalization group will be published elsewhere.) Similarly,
tremendous gains in efficiency would result from developing \SU N-symmetric
implementations of the density matrix renormalization group for treating
generic quantum chain models \cite{McCulloch2002,McCulloch2007}, or
generalizations of this approach for treating two-dimensional
tensor network models \cite{Singh2009}. 

For explicit calculations with models having \SU N symmetry, explicit
tables of \SU N Clebsch-Gordan coefficients are needed. Their
calculation is a problem of applied representation theory of
Lie groups that has been solved, in principle, long ago
\cite{Biedenharn1963,Baird1963,Baird1964a,Baird1964b,Baird1965}.
For example, for \SU 2 Racah\cite{Racah1942} has found an explicit
formula that gives the \CGC{}s for the direct product decomposition of
two arbitrary irreps $S$ and $S'$.  For \SU N, explicit \CGC{} formulas
exist for certain special cases, e.g.\ where $S'$ is the defining
representation \cite{Vilenkin1991,Vilenkin1992a,Vilenkin1992b}.
Moreover, symbolic packages such as the program ``Lie'' \cite{LiE} also
allow the computation of certain \CGC{}s, but rather have been conceived as a
general-purpose software for manipulating Lie algebras than a high-speed
implementation for calculating \CGC{}s.
However, for the general case no explicit \CGC{} formulas are known that would
constitute a generalization of Racah's results to arbitrary $N$, $S$ and $S'$.

The present paper describes a numerical solution to this problem, by
presenting an elementary but efficient algorithm (and a computer implementation thereof) for
producing explicit tables of \CGC{}s arising in the direct product
decomposition of two arbitrary \SU N irreps, for arbitrary $N$.
(Since \SU N and \SLC N have the same CGCs, our algorithm also applies to the
latter, but for definiteness we shall usually refer only to the former.)
Our work is addressed at a readership of physicists. Our algorithm uses only
elementary facts from \SU N representation theory, which we introduce and
summarize as needed, presuming only knowledge of \SU 2 representation theory
at a level conveyed in standard quantum mechanics textbooks. Previous attempts at
formulating an algorithm for calculating \SU N \CGC{}s are either not
sufficiently general for our purposes\cite{Williams1994,Rowe1997}, or require mathematical
methods\cite{Gliske2007} much more advanced than ours,
far beyond the scope of a standard physics education.

We begin in Sec.~\ref{sec:defineCGC} by formulating the problem
  to be solved in rather general terms. To set the scene for its
  solution, sections~\ref{sec:su2} to \ref{sec:raising-lowering} 
  summarize the various elements of \SU N representation theory
  (without proofs, since this is all textbook material).  First, in
  Sec.~\ref{sec:su2} we review the calculation of \SU 2 CGCs using a
  strategy that can readily be generalized to the case of \SU N. Then
  we proceed to \SU N representation theory and review in sections
  \ref{sec:algebra} to \ref{sec:raising-lowering} a scheme, due to
  Gelfand and Tsetlin (GT) \cite{Gelfand1950}, for labeling the
  generators of the corresponding Lie algebra \su N, its irreps and
  the states in each irrep. The GT-scheme is convenient for our
  purposes since it yields explicit matrix representations for any \SU
  N irrep (Eqs.~(\ref{eq:jmelement}) and (\ref{eq:jpelement}) below).
  With these in hand, we are finally in a position to formulate, in
  sections \ref{sec:decompose} to \ref{sec:algorithm}, our novel
  algorithm for computing \SU N CGCs: it is simply a suitably
  generalized version of the \SU 2 strategy of Sec.~\ref{sec:su2}.

  The main text is supplemented by several technical appendices.
  App.~\ref{app:young} reviews the relation between the GT-patterns
  used in the text and Young tableaux, with which physicists are
  perhaps somewhat more familiar.
  App.~\ref{app:Littlewood-Richardson} deals with the
  Littlewood-Richardson rule for determining which irreps
  $\mathbb{V}^{S''}$ occur in the decomposition $\mathbb{V}^{S}
  \otimes \mathbb{V}^{S'}$. App.~\ref{app:map} describes two
  algorithms, needed for indexing purposes, which map the labels of
  irreps and of carrier states, respectively, onto natural
  numbers.
Finally, App. \ref{app:source}, which is available in electronic form\cite{EPAPS},
gives the source code for our computer implementation, written in C++.
As a service to potential users, we have set up a web
site\cite{ClebschGordanWebSite} containing an interactive ``CGC-generator''.
It allows visitors to perform a number of tasks on input data of their own choice,
such as finding all irreps $S''$ occuring in the decomposition of $S \otimes S'$,
or finding the complete set of CGCs arising in the decomposition of $S \otimes S'$.
\section{Statement of the problem}
\label{sec:defineCGC}
To fix notation, let us state the problem we wish to solve
for a general matrix Lie group $\mathcal{G}$. (In subsequent sections,
we restrict attention to $\mathcal{G}$ = \SU N or \SLC N.)
Let $S$ be an irrep label that distinguishes different irreps of $\mathcal{G}$
of \SU N, and $d_S$ the dimension of irrep $S$. Let $\mathbb{V}^S =
\Span \{\ket{M}\}$ denote the carrier space for $S$, spanned by
$d_S$ carrier states $\ket{M}$,
where the label $M$ will be
understood to specify both the irrep $S$ and a particular state in its
carrier space. (This will be made explicit in subsequent sections.)
Note that, throughout this paper, we adopt the viewpoint of quantum mechanics,
where we consider only representations on \emph{complex} vector spaces.
Besides, a state is to be understood as a one-dimensional subspace, not a vector.
However, we pick a representative vector $\ket{M}$ of each such subspace
and subsequently treat a state as a vector. We assume the inner product of two such
normalized vectors $\ket{M}$ and $\ket{M'}$ to be given by
$\braket{M|M'} = \delta_{M,M'}$ unless noted otherwise.

The action of a group element $g \in \mathcal{G}$ can be
represented on $\mathbb{V}^S$ as a linear transformation 
\begin{equation}
  g: \ket{M} \to \sum_{M'} (U^S_g)_{M M'} \ket{M'} \; , 
\end{equation}
where the $U^S_g$ are $d_S \times d_S$ dimensional unitary matrices
respecting the group structure $U^S_{g_1} U^S_{g_2} = U^S_{g_1 g_2}$.

Now consider the direct product of two carrier spaces, $\mathbb{V}
\otimes \mathbb{V}' = \Span \{\ket{M \otimes M'}\}$, of
dimension $d_S \cdot d_{S'}$.  We are interested in its decomposition
into a direct sum of carrier spaces $\mathbb{V}^{S''}$ of irreps
$S''$,
\begin{equation}
\label{eq:directproductdecomposition}
\mathbb{V}^S \otimes \mathbb{V}^{S'}
= \bigoplus_{S''} \bigoplus^{N^{S''}_{S S'}}_{\alpha = 1} \mathbb{V}^{S'',\alpha}
\equiv \bigoplus_{S''} N^{S''}_{S S'} \mathbb{V}^{S''} .
\end{equation}
Here the integer $N^{S''}_{S S'} \ge 0$, called the \emph{outer multiplicity}
of $S''$, specifies the number of times the irrep
$S''$ occurs in this decomposition, and for a given $S''$, the outer
multiplicity index $\alpha = 1, \dots, N^{S''}_{S S'}$ distinguishes
multiple occurrences of $S''$. Correspondingly, let $\{\ket{M'', \alpha}\}$
be a basis for the direct sum decomposition, i.e.\
$\mathbb{V}^{S'', \alpha} = \Span \{\ket{M'',\alpha}\}$.
Carrier space dimensions add up according to $d_S \cdot d_{S'} = \sum_{S''}
N^{S''}_{S S'} d_{S''}$.

The decomposition (\ref{eq:directproductdecomposition}) implies that a
basis transformation $C$ can be found from the direct product basis to
the direct sum basis which block-diagonalizes the matrix
representations of all group elements (Ref.~\onlinecite{Cornwell1984a}, p.~100):
\begin{subequations}
\label{eq:blockdiagonalize}
\begin{align}
   C( U^S_g \otimes U^{S'}_g) C^\dagger = 
\begin{pmatrix} 
   U^{\tilde S_1}_g & & & \\
   & U^{\tilde S_2}_g & &  \\
   & & U^{\tilde S_3}_g & \\
   & & & \ddots   
\end{pmatrix} \; ,
\end{align}
where each $\tilde S_j$ is a shorthand for a certain $(S'',\alpha)$.

Since $\mathcal{G}$ is a matrix Lie group (\SU N or \SLC N), it
  is convenient to work with its associated Lie algebra $\mathfrak{g}$
  (\su N or \slc N). It is obtained by considering the infinitesimal
  action of $\mathcal{G}$ on $\mathbb{V}^{S}$, i.e.\ by taking
  derivatives of the group at the identity.
This derivative
acts on the direct product of two group representations according to
the product rule, so that the basis transformation $C$ could equally
be defined by the property that it block-diagonalizes the algebra
representation:
\begin{align}
   C( U^S_A \otimes \mathbb{I}^{S'} + \mathbb{I}^{S} \otimes U^{S'}_A) C^\dagger = 
\begin{pmatrix} 
   U^{\tilde S_1}_A & & & \\
   & U^{{\tilde S_2}}_A & &  \\
   & & U^{{\tilde S_3}}_A & \\
   & & & \ddots   
\end{pmatrix} \; . 
\end{align}
\end{subequations}
When projected
to the subspace $\mathbb{V}^{S'',\alpha}$ (denote the corresponding
projector by $P^{S'',\alpha}$), 
the action of the algebra in the direct product representation
can thus be written as
\begin{equation}
\label{eq:algebra-product-action}
C ( U_A^{S} \otimes \mathbb{I}^{S'} + \mathbb{I}^S \otimes 
U_A^{S'}) C^\dagger \quad \stackrel{P^{S'',\alpha}}{\longrightarrow}
\quad  U_A^{S'',\alpha} \; .
\end{equation}
Concretely, the basis transformation $C$ can be expressed in the form
\begin{align}
  \label{eq:defineCGC}
  \ket{M'', \alpha} = \sum_{M,M'} C^{M'', \alpha}_{M,M'} 
  \ket{M \otimes M'} \; ,
\end{align}
where the $C^{M'',\alpha}_{M,M'}$ are the Clebsch-Gordan-coefficients
  of present interest. They are understood to be defined only for
  $N^{S''}_{S,S'} \ne 0$, and express the carrier states of
  $\mathbb{V}^{S'',\alpha}$ in terms of linear combinations of product
  basis states from $\mathbb{V}^{S} \otimes \mathbb{V}^{S'}$. The
  \CGC{}s encode so-called \emph{selection rules}, in that $C^{M'',\alpha}_{MM'} \neq 0$
  only for a limited number of combinations of
  $M$, $M'$ and $M''$.

Since the CGCs are the entries of the unitary
matrix $C$, they satisfy the following orthonormality conditions:
\begin{subequations}
\label{eq:CGC-orthonormal}
\begin{align}
\label{eq:CGC-normalization}
\sum_{M,M'} C^{M'',\alpha}_{M,M'} (C^{\tilde M'',\tilde\alpha}_{M,M'})^\ast
&= \delta_{M'',\tilde M''} \delta_{\alpha,\tilde\alpha} \; , \\
\sum_{M'',\alpha} C^{M'',\alpha}_{M,M'} (C^{M'',\alpha}_{\tilde M,\tilde M'})^\ast
&= \delta_{M,\tilde M} \delta_{M',\tilde M'} \; . 
\end{align}
\end{subequations}
Actually, the $C^{M'',\alpha}_{M,M'}$ can always be chosen to be
  real, and we shall do so throughout.

The goal of the present work is to present (and implement on a
computer) an efficient algorithm for ${\cal G}$ = \SU N or
\SLC N which, for any specified $N$ and any specified irrep
labels $S$ and $S'$, produces explicit tables of all CGCs arising in
the direct product decomposition
\eqref{eq:directproductdecomposition}.
\section{Review of \SU 2 Clebsch-Gordan coefficients}
\label{sec:su2}
Before considering the general \SU N case, we first review a method for
calculating \SU 2 \CGC{}s.  While there are various ways to accomplish this
task, the particular approach presented below illustrates the general
strategy to be used for \SU N in later sections.
The discussion is structured as follows:
First, we recall the Lie algebra associated with \SU 2,
then its irreducible representations,
then move on to product representation decompositions,
and finally set up equations specifying the \CGC{}s.

The Lie algebra associated with \SU 2, denoted by \su 2,
consists of all real linear combinations of three basis elements,
$J_x$, $J_y$, and $J_z$, obeying the commutation relation $[J_x, J_y] = iJ_z$
(plus cyclic permutations of the indices). However,
it will be more convenient to deal with complex linear
combinations of these, which constitute the algebra \slc 2.
As a basis for the latter, it is common to choose three elements,
$J_+ = J_x + iJ_y$, $J_- = J_x - iJ_y$, and $J_z$,
obeying the following commutation relations:
\begin{subequations}
\label{eq:sl2-commutators}
\begin{align}{}
[J_z, J_\pm] &= \pm J_\pm, \\
[J_+, J_-] &= 2 J_z.
\end{align}
\end{subequations}
Each \su 2 irrep, and correspondingly, each \SU 2 irrep, can be uniquely
(up to an isomorphism) identified by a nonnegative half-integer,
$S = 0, 1/2, 1, \ldots$.  The carrier space $\mathbb{V}^S$ of such an irrep
has an orthonormal basis where the states, denoted by $\ket{S,m}$,
are labeled by a half-integer, $m = S, S-1, \ldots, -S$,
such that the action of $J_z$ and $J_\pm$ is given by
\begin{subequations}
\begin{align}
\label{eq:su2-jz-action} J_z\ket{S, m} &= m\ket{S, m}, \\
\label{eq:su2-jpm-action} J_\pm\ket{S, m} &= \sqrt{(S \pm m + 1)(S \mp m)}\ket{S, m \pm 1}.
\end{align}
\end{subequations}
\begin{figure}
\begin{center}
  \includegraphics{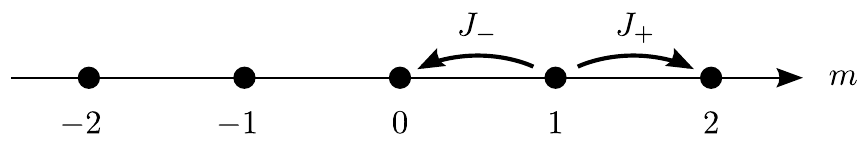}
\end{center}
\caption{\SU 2 weight diagram for $S=2$. Arrows show the action of $J_\pm$ on the state $\ket{S=2,m=1}$.}
\label{fig:su2}
\end{figure}
The $J_z$-eigenvalue $m$ will be called the $z$-\emph{weight} of the
state $\ket{S,m}$ (in anticipation of similar nomenclature to be used for
\SU N below). The action of $J_\pm$ can be visualized
in a so-called \emph{weight diagram},
which represents each carrier state $\ket{S,m}$
by a mark on an axis at the corresponding $m$-value.
For example, the carrier space of $S = 2$ is shown
in Fig.~\ref{fig:su2}. In anticipation of the generalization to \SU N,
we label basis states from now on by a composite index $M = (S, m)$,
which includes both the irrep label $S$ and the basis index $m$.

Each carrier space $\mathbb{V}^{S}$ contains a unique (up to normalization)
\emph{highest-weight state}, $\ket{H''}$, defined by the property that
\begin{equation}
\label{eq:SU2hw}
J_+ \ket{H} = 0 \;.
\end{equation}
For \su 2, it carries the labels $\ket{H} = \ket{S, m=S}$.

In the direct product decomposition of two \su 2 irreps $S$ and $S'$, the outer multiplicity $N^{S''}_{S,S'}$ in the notation of Eq.~\eqref{eq:directproductdecomposition} is given by:
\begin{equation}
\label{eq:SU2-outermultiplicities}
N^{S''}_{S,S'} = \begin{cases}
1 & \text{for } |S-S'| \le S'' \le S+S', \\
0 & \text{otherwise}.
\end{cases}
\end{equation}
Since $N^{S''}_{S,S'} \le 1$ for \su 2, we shall, throughout this section,
omit the index $\alpha$ appearing in Eq.~\eqref{eq:defineCGC}.
In particular, Eq.~\eqref{eq:defineCGC} now takes the form
\begin{equation}
\label{eq:su2-CGC}
\ket{M''} = \sum_{M,M'} C^{M''}_{M,M'} \ket{M \otimes M'},
\end{equation}
where the \CGC{}s $C^{M''}_{M,M'}$ satisfy the selection rule:
\begin{align}
\label{eq:SU2selectionrule-2}
m'' \neq  m + m'  \quad &\Longrightarrow \quad C^{M''}_{M,M'} = 0 \; .
\end{align}
It reflects the fact that $\ket{M''}$, 
$\ket{M}$ and $\ket{M'}$ are eigenstates of 
$J_z^{S''}$, $J_z^{S}$ and $J_z^{S'}$, respectively,
where the superscripts on $J_z$ indicate which carrier space
the respective operators act on. 

To obtain the \CGC{}s for given $S$ and $S'$ explicitly, we consider each $S''$
for which $N^{S''}_{S,S'} > 0$ separately.
Let us make the following ansatz for the expansion of $\ket{H''}$
in terms of product basis states:
\begin{equation}
\label{eq:defH-SU2}
\ket{H''} = \sum_{M,M'} C^{H''}_{M,M'} \ket{M \otimes M'},
\end{equation}
where $C^{H''}_{M,M'}$ are the \CGC{}s of $\ket{H''}$, and the sum runs only
over values of $m$ and $m'$ that satisfy the selection rule
\eqref{eq:SU2selectionrule-2}.
Inserting \eqref{eq:defH-SU2} into \eqref{eq:SU2hw}, we obtain
\begin{equation}
\label{eq:CGC-for-highest-weight}
\sum_{M,M'} C^{H''}_{M,M'} (J_+^S \otimes \mathbb{I}^{S'}
+ \mathbb{I}^{S} \otimes J_+^{S'}) \ket{M \otimes M'} = 0 \;.
\end{equation}
After evaluating the action of the raising operators on
$\ket{M \otimes M'}$ using Eq.~\eqref{eq:su2-jpm-action} and requiring
the coefficients in front of each state $\ket{M \otimes M'}$ to vanish
independently, we obtain a homogeneous linear system of equations.  We solve
for $C^{H''}_{M,M'}$ and fix a solution
by the normalization condition~\eqref{eq:CGC-normalization} and
by requiring $C^H_{M,M'}$ to be real and positive for
the largest value of $m$ for which $C^H_{M,M'}$ is nonzero.

The \CGC{}s of lower-weight states (i.e.\ states other than the highest-weight state) are found
by noting that
\begin{equation}
\label{eq:CGC-for-lower-weight}
\begin{aligned}
\ket{M''} = \ket{S'', m''} &= \mathcal{N} (J_-)^{S'' - m''} \ket{H''} \\
&= \mathcal{N} \sum_{M,M'} C^{H''}_{M,M'} (J_-^S \otimes \mathbb{I}^{S'} 
+ \mathbb{I}^{S} \otimes J_-^{S'})^{S''-m''} \ket{M \otimes M'}.
\end{aligned}
\end{equation}
($\mathcal{N} = \sqrt{(S''+m'')!/(S''-m'')!(2S'')!}$ is a normalization constant.)
The right-hand side of this equation is 
fully known from Eq.~\eqref{eq:su2-jpm-action}. By rewriting it 
into the form of Eq.~\eqref{eq:su2-CGC}, 
the desired $C^{M''}_{M,M'}$ can readily be identified.

For given $S''$, $S$ and $S'$ it is possible to write Eq.~\eqref{eq:CGC-for-lower-weight}
as a recursion relation relating \CGC{}s with different~$m''$~\cite{Sakurai1994}. 
Moreover, for \su 2, there exists a closed formula for $C^{M''}_{M,M'}$\cite{Racah1942}.
Nevertheless, for present purposes, the approach presented here is the most convenient
as its key steps can readily be generalized to calculate \su N Clebsch-Gordan coefficients.
The differences in comparison to \su 2 will lie in 
\begin{inparaenum}[(i)]
\item the more complex structure of raising and lowering operators,
\item the labeling schemes for irreps and states, and
\item the method for finding the irreps occurring in a product representation decomposition,
\end{inparaenum}
all of which we tackle in the following sections.
\section{The Lie algebra associated with \SU N}
\label{sec:algebra}
Instead of working with the group \SU N  itself, it
  will be more convenient for our purposes to consider its associated
  Lie algebra, \su N  \cite[ch.~13]{Cornwell1984b}. The
  latter consists of all traceless anti-Hermitian $n \times n$
matrices, while the ordinary commutator serves as its Lie
bracket. Most results obtained for representations of \su N carry over to \SU N
one-to-one, with the elements of the Lie algebra representing the
generators of the Lie group. Notably, the Clebsch-Gordan coefficients
of their representations are identical.

We begin by specifying a basis for the \su N algebra, in
order to illustrate its structure.
Let $E^{p,q}$ be the single-entry matrices, i.e.\
$E^{p,q}_{r,s} = \delta_{p,r}\delta_{q,s}$. A possible choice of basis
is given by the matrices $i(E^{k,l} + E^{l,k})$ and $E^{k,l}
- E^{l,k}$ for $1 \le k < l \le N$, and $i(E^{l,l} - E^{l+1,l+1})$
for $1 \leq l \le N-1$.
\su N is spanned by \emph{real}
linear combinations of these matrices.
Just as for \su 2 and \slc 2, however,
it will be convenient to work with a basis for \slc N.
To this end, define
for $1 \le l \le N-1$ the \emph{complex} linear combinations,
\begin{subequations}
\begin{align}
\label{eq:jzk}
J^{(l)}_z &=  \frac{1}{2}(E^{l,l} - E^{l+1,l+1}), \\
\label{eq:jpk}
J^{(l)}_+ &=  E^{l,l+1} , \\
\label{eq:jmk}
J^{(l)}_- &= E^{l+1,l} ,
\end{align}
\end{subequations}
which satisfy, for each $l$, the familiar \su 2
commutation relations of Eq.~\eqref{eq:sl2-commutators}:
\begin{subequations}
\begin{align}
\label{eq:zcomm}
\left[ J_z^{(l)}, J_\pm^{(l)} \right] &=  \pm J_\pm^{(l)}  , \\ 
\label{eq:pmcomm}
\left[ J_+^{(l)}, J_-^{(l)} \right] &=  2J_z^{(l)} . 
\end{align}
\end{subequations}
The $N-1$ matrices $J^{(l)}_z$
form a maximal set of mutually commuting matrices,
$[J^{(l)}_z, J^{(l')}_z] = 0$
(thus, the $i J^{(l)}_z$ span the Cartan subalgebra of \su N).
Thus, none of the $J^{(l)}_\pm$ commutes with all elements of this set,
or with all other $J^{(l')}_\pm$ operators.

The matrices $J^{(l)}_z$ and $J^{(l)}_\pm$ are not anti-Hermitian
and thus do not belong to \su N, 
but rather to \slc N.
However, it is sufficient to restrict our attention to
$J^{(l)}_\pm$ because, from these, we can recover an anti-Hermitian
basis using
\begin{subequations}
\begin{equation}
E^{p,q} = [J_-^{(p-1)}, [J_-^{(p-2)}, \ldots [J_-^{(q+1)}, J_-^{(q)}]] \ldots]
\quad\text{for}\quad p>q \;,
\end{equation}
\begin{equation}
E^{p,q} = [J_+^{(p)}, [J_+^{(p+1)}, \ldots [J_+^{(q-2)}, J_+^{(q-1)}]] \ldots]
\quad\text{for}\quad p<q \;.
\end{equation}
\end{subequations}
In other words, once we know representations for all $J^{(l)}_\pm$
on a given carrier space, the representations of all other elements of
both the algebras \slc N and \su N are also known.
For definiteness, we shall refer to \su N below,
although the constructions apply equally to \slc N.
\section{Labeling of irreps and states}
\label{sec:gelfand}
The \su N basis defined in the preceding section has a feature that
makes it particularly convenient for our purposes: if one also adopts
a specific labeling scheme, devised by Gelfand and Tsetlin (\GT{})
\cite{Gelfand1950}, for labeling \su N irreps and the basis states
of their carrier spaces,
these basis states are simultaneous
eigenstates of all the matrices $J^{(l)}_z$, and explicit formulas exist
for the matrix elements of the $J^{(l)}_\pm$ with respect to these
basis states.
The next three sections are devoted to summarizing the
GT labeling scheme without dwelling on its mathematical roots --
the mere knowledge of its rules is sufficient for our purposes. (The
  relation of the GT-scheme labeling scheme to a frequently-used
  alternative but equivalent labeling scheme, employing Young diagrams
  and Young tableaux, is summarized, for convenience,
in Appendix~\ref{app:young}.)

Up to equivalent representations, each \su N irrep can be identified
uniquely by a sequence of $N$ integers \cite{Biedenharn1968},
\begin{equation}
S = (m_{1,N}, \ldots, m_{N,N}),
\end{equation}
or $S = (m_{k,N})$ in short,
fulfilling $m_{k,N} \ge m_{k+1,N}$ for $1 \le k \le N-1$.
We shall call such a sequence an \emph{irrep weight} or \emph{\iweight},
in short.
The second index, $N$, identifies the algebra, \su N; the reasons for
displaying this index explicitly will become clear below. Two \iweight{}s
$S$ and $S'$ for which all components differ only by a
$k$-independent constant, i.e.\ $m'_{k,N} =
m_{k,N} + c$ with $c \in \mathbb{Z}$, designate the \emph{same} \su N
irrep. This fact can be used to bring any \iweight{} into a ``normalized''
form having $m_{N,N} = 0$, which will be assumed below, unless
otherwise specified.

\GAT{} exploited the fact that the carrier space of any \su N irrep
splits into disjoint carrier spaces of \su{N-1} irreps
to devise a labelling scheme with a very convenient property: 
It yields a remarkably simple rule for enumerating which
\su{N-1} irreps occur in the decomposition of $S = (m_{k,N})$,
namely all those with
\iweight{}s $(m_{1,N-1}, \ldots, m_{N-1,N-1})$ that satisfy the condition
$m_{k,N} \ge m_{k,N-1} \ge m_{k+1,N}$ for $1 \le k \le N-1$. Note that,
here, it is crucial \emph{not} to set $m_{N-1,N-1} = 0$ so that we can
distinguish between multiple occurrences of the same \su{N-1} irrep.

Recursively, the carrier spaces of \su{N-1} irreps give rise to
\su{N-2} irreps and so on, down to \su 1, the carrier spaces of which
are one-dimensional. This sequence of decompositions can be exploited
to label the basis states $\ket{M}$ of a given \su N irrep
$S=(m_{k,N})$ using so-called \emph{Gelfand-Tsetlin patterns}
(GT-patterns). These are triangular arrangements of integers, to
be denoted by $M = (m_{k,l})$, with the structure
\begin{equation}
\label{eq:gtpattern}
M =  \begin{pmatrix}
\multicolumn{2}{c}{m_{1,N}} & \multicolumn{2}{c}{m_{2,N}} &
\multicolumn{2}{c}{\ldots} & \multicolumn{2}{c}{m_{N,N}} \\
& \multicolumn{2}{c}{m_{1,N-1}} & \multicolumn{2}{c}{\ldots}
& \multicolumn{2}{c}{m_{N-1,N-1}} & \\
&& \ddots &&& \reflectbox{\(\ddots\)} && \\
&& \multicolumn{2}{r}{m_{1,2}} & \multicolumn{2}{l}{m_{2,2}} && \\
&&& \multicolumn{2}{c}{m_{1,1}} &&&
\end{pmatrix},
\end{equation}
i.e.\ the first index labels diagonals from left to right,
and the second index labels rows from bottom to top.
The top row contains the \iweight{} $(m_{k,N})$ that specifies the
irrep, and the entries of lower rows are subject to the so-called
\emph{betweenness condition},
\begin{align}
\label{eq:betweenness}
m_{k,l} &\geq m_{k,l-1} \geq m_{k+1,l} & (1 \leq k < l \leq N).
\end{align}
The dimension of an irrep $S=(m_{k,N})$ is equal to the number of valid
\GTP{}s having $S$ as their top row. There exists a
convenient formula for this number:
\begin{equation}
\label{eq:dimension}
\dim (S) = \prod_{1 \leq k < k' \leq N} \left( 1 + \frac{m_{k,N} - m_{k',N}}{k' - k} \right).
\end{equation}
Note that the \SU 2 basis state conventionally labeled as $\ket{j,m}$ corresponds to the \GTP{}
{\tiny$\begin{pmatrix} 2j \quad & \quad 0 \\ \multicolumn{2}{c}{j-m} \end{pmatrix}$}, and the above formula reduces to
$\dim(j) = 2j+1$.

To obtain a complete description of \SU N irreps, we need to specify
how the Lie algebra \su N acts on states labeled by Gelfand-Tsetlin
patterns. The following two sections are devoted to this task,
section~\ref{sec:weights} with $J^{(l)}_z$
and section~\ref{sec:raising-lowering} dealing with $J^{(l)}_\pm$.
\section{Weights and weight diagrams}
\label{sec:weights}

A very convenient property of the GT-labeling scheme is
that every state $\ket{M}$ is a  simultaneous 
eigenstate of all $J^{(l)}_z$ generators,
\begin{align}
J^{(l)}_z \ket{M} &= \lambda^M_l \ket{M}, & (1 \le l \le N-1),
\end{align}
with eigenvalues
\begin{align}
\label{eq:jzelement}
\lambda^M_l &= \sigma^M_l - \frac{1}{2} (\sigma^M_{l+1} + \sigma^M_{l-1}) & (1 \leq l \leq N-1),
\end{align}
where the \emph{row sum} $\sigma^M_l = \sum_{k=1}^l m_{k,l}$ denotes the sum over all
entries of row $l$ of \GTP{} $M$ ($\sigma^M_0 = 0$ by convention).  We
shall call the sequence of $N-1$ $J^{(l)}_z$ eigenvalues the
  \emph{$z$-weight} of the state $\ket{M}$, and denote it by $W_z(M)
  = (\lambda_1^M, \ldots, \lambda_{N-1}^M)$. The
  \emph{$z$-weight} of $\ket{M}$ is a straightforward generalization
  of the quantum number $m$ in quantum angular momentum. 

  As will be elaborated below, the notion of weights of states is useful for
  elucidating the structure of carrier spaces of \su N irreps, and in
  particular for visualizing the action of raising and lowering
  operators. The above way of introducing weights is, however, not
  unique. We shall often find it convenient to employ an alternative
  definition of the weight of states, which has the convenient property that it
  always yields nonnegative \emph{integer} elements (in contrast to $W_z (M)$). This
alternative weight, to be called \emph{pattern weight} or \emph{\pweight},
 and denoted by $W(M)$, is
defined to be a sequence of $N$ integers, $W(M) = (w_1^M, \ldots,
w_N^M)$, where
\begin{align}
w^M_l &= \sigma^M_l - \sigma^M_{l-1} & (1 \leq l \leq N)
\end{align}
is the difference between summing up rows $l$ and $l-1$ of the
  GT-pattern $M$.  Note that the number of \emph{independent}
elements of $W(M)$ is the same as that of $W_z(M)$, namely $N-1$,
since the $w_l^M$ satisfy the relation $\sum_{l=1}^{N} w^M_l =
\sigma^M_N$. The two types of weights are directly related to each
other: via Eq.~\eqref{eq:jzelement}, we obtain $\lambda^M_l = (w^M_l -
w^M_{l+1})/2$.
For definiteness, we will mostly refer to \pweight{}s below
(noting here that most statements involving \pweight{}s can be translated
into equivalent statements involving $z$-weights).

At this point, the first of several fundamental differences between
\su 2 and \su N with $N \geq 3$ appears. While for \su 2, there always
exists exactly one state with a given \pweight{}, this is not the case for
\su N in general; for $N \geq 3$, several linearly independent states
in the carrier space can have the same \pweight{}.
Indeed, two states have the same \pweight{}, 
$W(M) = W(M')$, if and only if they
have the  same set of row sums ($\sigma_l^M = \sigma_l^{M'}$  
for $1 \le l \le N-1$)  (i.e.\ they differ only 
in the way in which the ``weight'' of the row sums
is distributed among the entries of each row). 
 For a given \pweight{} $W$,
the number of states $\ket{M}$ having the same \pweight{}, $W(M) = W$, is
called the \emph{inner multiplicity} of that \pweight{}, to be denoted by
$I(W)$. Consequently, \pweight{}s or $z$-weights are not suited for uniquely labeling
states of a carrier space (which is why \GTP{}s are used for this purpose).

\begin{figure}
\begin{center}
\includegraphics{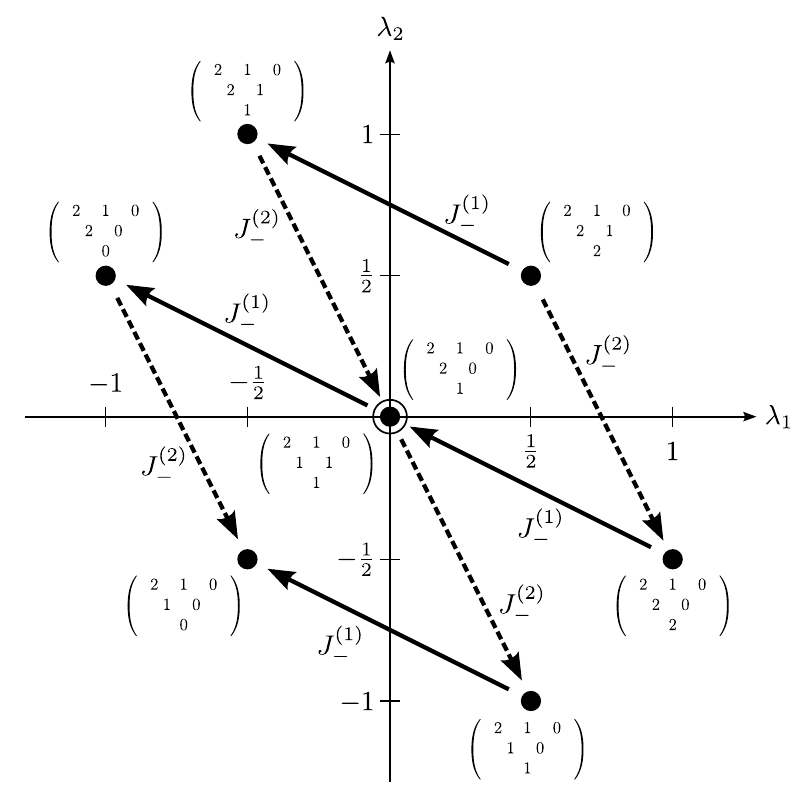}
\end{center}
\caption{Weight diagram of the \su 3 irrep $(2,1,0)$. Each dot
  represents a $z$-weight; we also indicate the \GTP{}s
of the corresponding states. The double circle around $(0,0)$ indicates that
  there are two states with this weight.
The solid and dashed arrows represent the action of $J^{(1)}_-$
and $J^{(2)}_-$, respectively.
($J^{(l)}_+$ could be represented by arrows
  pointing in directions opposite to those of $J^{(l)}_-$.)
  Note that both $J^{(1)}_-$ acting
  on {\tiny\protect\gt{2}{1}{0}{2}{0}{2}} and $J^{(2)}_-$ acting on
  {\tiny\protect\gt{2}{1}{0}{2}{1}{1}} produce linear combinations of
  {\tiny\protect\gt{2}{1}{0}{2}{0}{1}} and
  {\tiny\protect\gt{2}{1}{0}{1}{1}{1}}, albeit different ones.
(In the literature it is not uncommon to choose
a different \su 3 basis that renders this
  weight diagram more symmetric.)}
\label{fig:su3}
\end{figure}

$z$-weights nevertheless do provide a convenient way to visualize the
carrier space of an \su N irrep. To this end, consider $W_z(M) =
(\lambda^M_1, \ldots, \lambda^M_{N-1})$ as a vector in
$(N-1)$-dimensional space and, for each state, mark 
the endpoint of its weight vector in an $(N-1)$-dimensional
lattice. The resulting diagram is called a \emph{weight diagram}. For
the \su 2 irrep $j$, weight diagrams consist of a coordinate axis with
markings at $-j, -j + 1, \ldots, j$ (see Fig.~\ref{fig:su2});
for \su 3, weight diagrams are
two-dimensional (see Fig.~\ref{fig:su3}); for $N \geq 4$, weight
diagrams cannot be readily drawn on paper because the corresponding
lattices have more than two dimensions.

Note that, in Fig.~\ref{fig:su3}, the
$z$-weight $W_z = (0,0)$ has inner multiplicity two,
since the two states {\tiny\gt{2}{1}{0}{2}{0}{1}} and {\tiny\gt{2}{1}{0}{1}{1}{1}}
have the same row sums.
\section{Raising and lowering operators}
\label{sec:raising-lowering}
Weight diagrams are also very convenient for visualizing the action of
the raising and lowering operators $J^{(l)}_\pm$.
The action of $J^{(l)}_\pm$ on a given state $\ket{M}$
produces a linear combination of all states of the form $\ket{M \pm
  M^{k,l}}$ with arbitrary $k$, where this notation
implies element-wise addition and subtraction of the single-entry
pattern $M^{k,l}$ having $1$ at position $k,l$ and zeros elsewhere,
\begin{align}
\label{eq:single-entry-pattern}
M^{k,l} =  \begin{pmatrix}
\multicolumn{2}{c}{0} & \multicolumn{2}{c}{0} &
\multicolumn{2}{c}{\ldots} & \multicolumn{2}{c}{0} \\
& \multicolumn{2}{c}{0} & \multicolumn{2}{c}{\ldots}
& \multicolumn{2}{c}{0} & \\
&& \ddots &&1_{k,l}& \reflectbox{\(\ddots\)} && \\
&& \multicolumn{2}{r}{0} &  \multicolumn{2}{l}{\phantom{.} \quad 0} && \\
&&& \multicolumn{2}{c}{0} &&&
\end{pmatrix}.
\end{align}
(Note that $M^{k,l}$ on its own is not a valid GT-pattern.)
Thus the resulting patterns differ from $M$ only in row $l$.  
All states $\ket{M \pm M^{k,l}}$ that are generated in this fashion 
have the same row sums, $z$-weights, and \pweight{}s (independent of $k$),
\begin{subequations}
\label{eq:weight-shift}
\begin{align}
\label{eq:shifted-z-weight} W_z(M \pm M^{k,l}) &= (\lambda^M_1, \ldots, \lambda^M_{l-2}, \lambda^M_{l-1} \mp 1/2, \lambda^M_{l} \pm 1, \lambda^M_{l+1} \mp 1/2, \lambda^M_{l+2}, \ldots, \lambda^M_{N-1}), \\
\label{eq:shifted-gt-weight} W(M\pm M^{k,l}) &= (w^M_1, \ldots, w^M_{l-1}, w^M_l \pm 1, w^M_{l+1} \mp 1, w^M_{l+2}, \ldots, w^M_N),
\end{align}
\end{subequations}
unless states with this weight do not exist, in
which case the result vanishes.

The weight-shifting action of lowering operators is illustrated in
Fig.~\ref{fig:su3} for the weight diagram of the \su 3 irrep
$S=(2,1,0)$.  Since the weight diagram is two-dimensional, there are
two lowering operators, $J^{(1)}_-$ and $J^{(2)}_-$, which shift in
different directions (indicated by solid/dashed lines). ($J^{(l)}_+$ produces a shift in
the opposite direction of $J^{(l)}_-$.) Note that there are two different ``paths''
to reach the $z$-weight (0,0) from the $z$-weight $(\frac{1}{2},
\frac{1}{2})$, namely via either $J^{(1)}_- J^{(2)}_-$ or $J^{(2)}_-
J^{(1)}_-$.  Since $J^{(1)}_-$ and $J^{(2)}_-$ do not commute, these
paths are inequivalent; indeed, they produce two different linear
combinations of the two states with $z$-weight $(0,0)$.
More generally, the fact that inner multiplicities larger than 1 arise
for \su N representations with $N > 2$ is a direct consequence of
the fact that there are, in general, several different ways
of reaching one state from another via a chain of raising and 
lowering operators, and that these ways are not equivalent,
because $J^{(l)}_\pm$ and $J^{(l')}_\pm$ do not commute for $l \neq l'$.

Very conveniently, closed expressions have been found
by Gelfand and Tsetlin \cite{Gelfand1950}
for the matrix elements of all raising and lowering operators
with respect to the basis of GT-patterns.
Explicitly, the only nonzero matrix elements of
$J^{(l)}_-$ are given, for any $ 1 \le k \le l \le N-1$,
by \cite[p.~280]{Barut1986}:
\begin{equation}
\label{eq:jmelement}
\braket{M - M^{k,l} | J_-^{(l)} | M} = \left( -\frac{
\prod\limits_{k'=1}^{l+1} (m_{k',l+1} - m_{k,l} + k - k' + 1)
\prod\limits_{k'=1}^{l-1} (m_{k',l-1} - m_{k,l} + k - k')
}{
\prod\limits_{\substack{k'=1 \\ k' \neq k}}^{l} (m_{k',l} - m_{k,l} + k - k' + 1)
(m_{k',l} - m_{k,l} + k - k')
} \right)^{\frac{1}{2}} .
\end{equation}
These matrix elements are real and nonnegative, and the right-hand side vanishes if $M
- M^{k,l}$ is not a valid pattern. As $J^{(l)}_+$ is the Hermitian
transpose of $J^{(l)}_-$, we can obtain its nonzero matrix elements by taking
the complex conjugate of the preceding formula and replacing $\ket{M}$
by $\ket{M + M^{k,l}}$:
\begin{equation}
\label{eq:jpelement}
\braket{M + M^{k,l} | J_+^{(l)} | M} = \left( -\frac{
\prod\limits_{k'=1}^{l+1} (m_{k',l+1} - m_{k,l} + k - k')
\prod\limits_{k'=1}^{l-1} (m_{k',l-1} - m_{k,l} + k - k' - 1)
}{
\prod\limits_{\substack{k'=1 \\ k'\neq k}}^{l} (m_{k',l} - m_{k,l} + k - k')
(m_{k',l} - m_{k,l} + k - k' - 1)
} \right)^{\frac{1}{2}} .
\end{equation}
These formulae generalize Eq.~\eqref{eq:su2-jpm-action} to \su N.

Each irrep has a unique state $\ket{H}$, called its
\emph{highest-weight state}, that is annihilated by all $N-1$ raising
operators
\begin{align}
\label{eq:hwdef}
J^{(l)}_+ \ket{H} &= 0 & (1 \leq l \leq N-1).
\end{align}
Since $\ket{H}$ is a unique state, the inner multiplicity of its \pweight{}
$W(H)$ is one, and the irrep can be identified by specifying $W(H)$.
Our labeling scheme indeed exploits this fact: the \iweight{} of an irrep
is equal to the \pweight{} of its highest-weight state $\ket{H}$, i.e.\ $S = W(H)$.
Conveniently, the \GTP{} $H = (h_{k,l})$ has the highest
possible entries fulfilling Eq.~\eqref{eq:betweenness},
i.e.\ $h_{k,l} = h_{k,N}$ for $1 \le k \le l \le N-1$
(all entries on the $k$-th diagonal are equal to $m_{k,N}$).

This concludes our exposition of those elements of \SU N representation
theory in the GT-scheme that are needed
in this work. In the following sections we discuss the decomposition of direct product representations and the calculation of the associated \CGC{}s.
The specific details of the strategy described below are, to the best of our knowledge, original.
\section{Product representation decompositions}
\label{sec:decompose}
The product of two irreps, say $S \otimes S'$, is, in general,
reducible to a sum of irreps
(Eq.~\eqref{eq:directproductdecomposition}).
While it is well-known for \su 2 which
irreps occur in such a decomposition
(see Eq.~\eqref{eq:SU2-outermultiplicities}),
the corresponding result for \su N relies on a relatively simple but
hard to prove method based on the
\emph{Littlewood-Richardson rule}~\cite{Leeuwen2007}.
This method involves writing down all possible \GTP{}s for
the irrep $S$ and using each of these to construct, starting from
$S'$, a new irrep $S''$. As the outcome of this method is the same when
interchanging $S$ and $S'$, it is preferable to take the irrep with
the smaller dimension of the two as $S$.

For given irreps $S = (m_{k,N})$ and $S' = (m'_{k,N})$,
and a particular \GTP{} $M = (m_{k,l})$ associated with $S$,
let us introduce some auxiliary notation.
For $l = 1, \ldots, N$ and $k = 1,
\ldots, l$, we set $b_{k,l} = m_{k,l} - m_{k,l-1}$ (where $m_{k,l} \equiv 0$
if $k>l$, for ease of notation) and $B_{k,l} = m'_{l,N} + \sum_{k'=1}^{k} b_{k,l}$
(note that here, $m'_{k,l}$ carries a prime, while $b_{k,l}$ does not). Then,
the irrep $S'' = (m''_{k,N}) \equiv (B_{k,k})$ occurs
in the decomposition of $S \otimes S'$ if and only if
\begin{equation}
\label{eq:condition}
B_{k-1,1} \ge B_{k-1,2} \ge \cdots \ge
B_{k-1,l-1} \ge B_{k,l} \ge B_{k,l+1} \geq \cdots \geq B_{k,N} \quad
\text{for all} \quad 1 \leq k \leq l \leq N \;.
\end{equation}
(We emphasize that this condition must hold for \emph{each} value
of $k$ and $l$.) By checking whether \eqref{eq:condition} holds for
all \GTP{}s associated with $S$, all $S''$ in the decomposition of
$S \otimes S'$ can be identified.

There exists a more efficient way to validate
Eq.~\eqref{eq:condition} than to check each value of $k$ and $l$ independently.
For a given \GTP{} $M = (m_{k,l})$ associated with $S$,
proceed as follows:
\begin{enumerate}
\item Initialize $(t_1, \ldots, t_N) = (m'_{1,N}, \ldots, m'_{N,N})$
by the \iweight{} of $S'$.
\item Step through the pattern $M$ along the diagonals
from top to bottom and from left to right, i.e.\ in the order $m_{1,N}$,
$m_{1,N-1}$, $\ldots$, $m_{1,1}$, $m_{2,N}$, $m_{2,N-1}$, $\ldots$, $m_{2,2}$,
$\ldots$, $m_{N,N}$.
\item At each position, say $m_{k,l}$, replace $t_l$ by $t_l + b_{k,l}$.
\item If $l > 1$, check whether $t_{l-1} \geq t_l$. If this condition is
violated, discard this \GTP{}, construct the next one, and commence
again from step~1.
\item If we reach the end of the pattern $M$,
the current value of $(t_1, \ldots, t_N)$
specifies the weight of an irrep $S''$ that occurs
in the decomposition of $S \otimes S'$.
\end{enumerate}
For $N>2$, this procedure in general can produce several occurrences of the
same irrep $S''$. The number of such occurrences, denoted by $N^{S''}_{S S'}$
in Eq.~\eqref{eq:directproductdecomposition}, is the outer multiplicity of $S''$.
(For \SU 2, the outer multiplicity is either 0 or 1.)

Let us illustrate this procudure by an example (individual steps are
shown in Table~\ref{tab:decompose}):
\begin{subequations}
\begin{align}
\label{eq:directdecomp-su3-explicit}
  (2,1,0) \otimes (2,1,0) &= (4,2,0) \oplus (3,3,0) \oplus (4,1,1) \oplus  (3,2,1) \oplus (3,2,1) \oplus (2,2,2) \\
  &= (4,2,0) \oplus (3,3,0) \oplus (3,0,0) \oplus (2,1,0) \oplus (2,1,0) \oplus (0,0,0)
\end{align}
\end{subequations}
(For the second line, we adopted ''normalized'' \iweight{}s with
  $m^{\phantom{\dagger}}_{N,N} = 0$.)
To check that the
dimensions are correct, use Eq.\eqref{eq:dimension} to verify the
dimensions of the irreps in this equation are $8 \times 8 = 27 + 10 +
10 + 8 + 8 + 1$, respectively.
\begin{table}
\begin{center}
\begin{sideways}
\begin{tabular}{c|c||c|c|c|c|c|c|c||c}
  \vphantom{\fbox{\parbox{2.5cm}{all possible \\ {\tiny\gt{m_{1,3}}{m_{2,3}}{m_{3,3}}{m_{1,2}}{m_{2,2}}{m_{1,1}}}}}}
  \parbox{2.5cm}{all possible \\ {\tiny\gt{m_{1,3}}{m_{2,3}}{m_{3,3}}{m_{1,2}}{m_{2,2}}{m_{1,1}}}}
  & \parbox{2.5cm}{corresponding \\ {\tiny\gt{b_{1,3}}{b_{2,3}}{b_{3,3}}{b_{1,2}}{b_{2,2}}{b_{1,1}}}}
  & \parbox{2.5cm}{$\text{initial value} = (m'_{1,3}, m'_{2,3}, m'_{3,3})$}
  & \parbox{1.7cm}{$\text{previous} + (0, 0, b_{1,3})$}
  & \parbox{1.7cm}{$\text{previous} + (0, b_{1,2}, 0)$}
  & \parbox{1.7cm}{$\text{previous} + (b_{1,1}, 0, 0)$}
  & \parbox{1.7cm}{$\text{previous} + (0, 0, b_{2,3})$}
  & \parbox{1.7cm}{$\text{previous} + (0, b_{2,2}, 0)$}
  & \parbox{1.7cm}{$\text{previous} + (0, 0, b_{3,3})$}
  & final irrep \\
  \hline
  \vphantom{\fbox{\small\gt{2}{1}{0}{2}{1}{2}}} {\small\gt{2}{1}{0}{2}{1}{2}}
  & {\small\gt{0}{0}{0}{0}{1}{2}}
  & $(2,1,0)$ & $(2,1,0)$ & $(2,1,0)$ & $(4,1,0)$ & $(4,1,0)$ & $(4,2,0)$ & $(4,2,0)$ & $(4,2,0)$ \\
  \vphantom{\fbox{\small\gt{2}{1}{0}{2}{1}{1}}} {\small\gt{2}{1}{0}{2}{1}{1}} &
  {\small\gt{0}{0}{0}{1}{1}{1}}
  & $(2,1,0)$ & $(2,1,0)$ & $(2,2,0)$ & $(3,2,0)$ & $(3,2,0)$ & $(3,3,0)$ & $(3,3,0)$ & $(3,3,0)$ \\
  \vphantom{\fbox{\small\gt{2}{1}{0}{2}{1}{1}}} {\small\gt{2}{1}{0}{2}{0}{2}}
  & {\small\gt{0}{1}{0}{0}{0}{2}}
  & $(2,1,0)$ & $(2,1,0)$ & $(2,1,0)$ & $(4,1,0)$ & $(4,1,1)$ & $(4,1,1)$ & $(4,1,1)$ & $(4,1,1) = (3,0,0)$ \\
  \vphantom{\fbox{\small\gt{2}{1}{0}{2}{1}{1}}} {\small\gt{2}{1}{0}{2}{0}{1}}
  & {\small\gt{0}{1}{0}{1}{0}{1}}
  & $(2,1,0)$ & $(2,1,0)$ & $(2,2,0)$ & $(3,2,0)$ & $(3,2,1)$ & $(3,2,1)$ & $(3,2,1)$ & $(3,2,1) = (2,1,0)$ \\
  \vphantom{\fbox{\small\gt{2}{1}{0}{2}{1}{1}}} {\small\gt{2}{1}{0}{2}{0}{0}}
  & {\small\gt{0}{1}{0}{2}{0}{0}}
  & $(2,1,0)$ & $(2,1,0)$ & $(2,3,0)$ & & & & & discarded \\
  \vphantom{\fbox{\small\gt{2}{1}{0}{2}{1}{1}}} {\small\gt{2}{1}{0}{1}{1}{1}}
  & {\small\gt{1}{0}{0}{0}{1}{1}}
  & $(2,1,0)$ & $(2,1,1)$ & $(2,1,1)$ & $(3,1,1)$ & $(3,1,1)$ & $(3,2,1)$ & $(3,2,1)$ & $(3,2,1) = (2,1,0)$ \\
  \vphantom{\fbox{\small\gt{2}{1}{0}{2}{1}{1}}} {\small\gt{2}{1}{0}{1}{0}{1}}
  & {\small\gt{1}{1}{0}{0}{0}{1}}
  & $(2,1,0)$ & $(2,1,1)$ & $(2,1,1)$ & $(3,1,1)$ & $(3,1,2)$ & & & discarded \\
  \vphantom{\fbox{\small\gt{2}{1}{0}{2}{1}{1}}} {\small\gt{2}{1}{0}{1}{0}{0}}
  & {\small\gt{1}{1}{0}{1}{0}{0}}
  & $(2,1,0)$ & $(2,1,1)$ & $(2,2,1)$ & $(2,2,1)$ & $(2,2,2)$ & $(2,2,2)$ & $(2,2,2)$ & $(2,2,2) = (0,0,0)$ \\
\end{tabular}
\end{sideways}
\end{center}
\caption{Application of the Littlewood-Richardson rule according to the steps of Section~\ref{sec:decompose},
for the decomposition of $S \otimes S' = (2,1,0) \otimes (2,1,0)$.
}
\label{tab:decompose}
\end{table}
Note that the irrep $(2,1,0)$ occurs twice in the
decomposition, in other words, its outer multiplicity is 2.
\section{Selection rule for \SU N Clebsch-Gordan coefficients}
The fact that all states labeled by \GTP{}s are eigenstates of $J^{(l)}_z$ operators implies
a selection rule for \SU N \CGC{}s.
Explicitly, let us consider a state $\ket{M''}$ occurring in a
decomposition of a product representation.
On the one hand, we have
\begin{subequations}
\begin{equation}
J^{(l)}_z \ket{M'',\alpha} = \lambda^{M'',\alpha}_l \ket{M'',\alpha},
\end{equation}
and on the other hand,
by Eqs.~\eqref{eq:algebra-product-action} and~\eqref{eq:defineCGC},
\begin{equation}
  J^{(l)}_z \ket{M'',\alpha} = \sum_{M,M'} C^{M'',\alpha}_{M,M'} 
  (J^{(l),S}_z \otimes \mathbb{I}^{S'} + \mathbb{I}^{S} \otimes J^{(l),S'}_z)
  \ket{M \otimes M'}
  = \sum_{M,M'}  C^{M'',\alpha}_{M,M'} (\lambda^{M}_l + \lambda^{M'}_l) \ket{M \otimes M'} \; . 
\end{equation}
\end{subequations}
These equations can only be fulfilled if $C^{M'',\alpha}_{M,M'}$ vanishes
whenever $\lambda^{M'',\alpha}_l \ne \lambda^{M}_l + \lambda^{M'}_l$ for any $l$.
Defining an element-wise addition on weights, we write, in short:
\begin{equation}
\label{eq:sun-selection-rule}
W_z(M'') \ne W_z(M) + W_z(M') \Longrightarrow C^{M'',\alpha}_{M,M'} = 0.
\end{equation}
This equation (or a transcription thereof involving \pweight{}s)
represents the generalization of Eq.~\eqref{eq:SU2selectionrule-2} to \su N.
\section{Clebsch-Gordan coefficients of highest-weight states}
\label{sec:hwstate}
After determining which kinds of irreps $S''$ appear in the decomposition of a
product representation, we are ready to construct their Clebsch-Gordan
coefficients. For each $S''$, we start by finding
the CGCs of its highest-weight state, $\ket{H'',\alpha}$, as defined in Eq.~\eqref{eq:hwdef}.
The index $\alpha = 1, \ldots, N^{S''}_{S,S'}$ distinguishes between the instances of irreps with outer multiplicity.
Nevertheless, we determine the CGCs of $\ket{H'',\alpha}$ with given $S''$ for all values of $\alpha$ in a single run.

For this purpose,
we make an ansatz of the form \eqref{eq:defineCGC}
for the highest-weight state (compare Eq.~\eqref{eq:defH-SU2}),
\begin{equation}
\label{eq:hwexp}
\ket{H'',\alpha} = \smashoperator{\sum_{\substack{M, M' \\ W(M) + W(M') = W(H'',\alpha)}}} C^{H'',\alpha}_{M,M'} \ket{M \otimes M'},
\end{equation}
with \CGC{}s $C^{H'',\alpha}_{M,M'}$, where the sum is restricted to
those combinations of states $\ket{M \otimes M'}$ that respect the
selection rule \eqref{eq:sun-selection-rule}.
Now insert Eq.~\eqref{eq:hwexp} into Eq.\eqref{eq:hwdef}
to obtain (compare Eq.~\eqref{eq:CGC-for-highest-weight}),
\begin{align}
\label{eq:hw-clebsches}
\smashoperator{\sum_{\substack{M, M' \\ W(M) + W(M') = W(H'',\alpha)}}} C^{H'',\alpha}_{M,M'} 
(J^{(l),S}_+ \otimes \mathbb{I}^{S'} + \mathbb{I}^{S} \otimes J^{(l),S'}_+) \ket{M \otimes M'} &= 0, & (1 \le l \le N-1).
\end{align}
After evaluating the action of the raising operators on the
product basis states via Eq.~\eqref{eq:jpelement},
we obtain a homogeneous linear system of equations in the \CGC{}s $C^{H'',\alpha}_{M,M'}$.
\label{sec:outer-multiplicity}
%
%
It has $N^{S''}_{S,S'}$ linearly independent
solutions, one for each value of $\alpha$. Thus,
an outer multiplicity larger than~1 leads to an ambiguity among the \CGC{}s
of the highest-weight states of all irreps of the same kind $S''$:
a unitary transformation
$\ket{H,\alpha} \to \sum_{\alpha'} U_{\alpha,\alpha'} \ket{H,\alpha'}$ among
the highest-weight states will produce different,
but equally acceptable highest-weight \CGC{}s $C^{H'',\alpha}_{M,M'}$.
The full set of CGCs of the irreps $S''$ will change accordingly, too.
For some applications, there is no need to uniquely resolve this ambiguity.
For applications where it must be resolved, we will adopt the following convention, 
suggested by G.~Zar\'and \cite{Zarand2009}:
Write down the
  independent solutions in the form of a matrix with elements
  $C^{H'',\alpha}_{MM'}$, where $\alpha = 1, \dots, N^{S''}_{S,S'}$ serves as row index and $(M,M') = 1, \ldots, I(H)$
  as composite column index (where $I(H)$ is the inner
  multiplicity of $W(H)$ in the product representation). Then use
  Gaussian elimination to bring this matrix into a normal form, namely
  the reduced row echelon form,
\begin{equation}
\begin{pmatrix}
\cdot & \cdots & \cdot \\
\vdots & C^{H'',\alpha}_{M,M'} & \vdots \\
\cdot & \cdots & \cdot \\
\end{pmatrix} 
\quad \to \quad
\left( \begin{array}{ccccccccccccccc}
0 & \cdots & 0 & + & 0 & \cdots & 0 & 0 & 0 & \cdots & 0 & 0 & \ast & \cdots & \ast \\
0 & \cdots & 0 & 0 & 0 & \cdots & 0 & + & 0 & \cdots & 0 & 0 & \ast & \cdots & \ast \\
0 & \cdots & 0 & 0 & 0 & \cdots & 0 & 0 & 0 & \cdots & 0 & 0 & \ast & \cdots & \ast \\
\vdots & \ddots & \vdots & \vdots & \vdots & \ddots & \vdots & \vdots & \vdots & \ddots & \vdots & \vdots & \vdots & \ddots & \vdots \\
0 & \cdots & 0 & 0 & 0 & \cdots & 0 & 0 & 0 & \cdots & 0 & 0 & \ast & \cdots & \ast \\
0 & \cdots & 0 & 0 & 0 & \cdots & 0 & 0 & 0 & \cdots & 0 & + & \ast & \cdots & \ast \\
\end{array} \right),
\end{equation}
where $+$ and $\ast$ denote positive and arbitrary matrix elements,
respectively. This normal form is the same for all equivalent
matrices.
To obtain orthonormal highest-weight states, we then do a Gram-Schmidt
orthonormalization of the rows of the resulting matrix from top to bottom.
This procedure uniquely specifies the
\CGC{}s for the highest-weight states.

As an aside, we note that the abovementioned ambiguity does not arise
for the case of $S' = (1, 0, \ldots, 0)$ (the defining
representation of \su N) and arbitrary $S$, since then all outer multiplicites are either zero
or one, i.e.\ then $N_{S,S'}^{S''} = 0 \;\text{or}\; 1$.
(However, there would still be a sign ambiguity for the \CGC{}s,
and the above procedure constitutes one way of fixing it.)
We note that for this case, explicit formulas for \SU N \CGC{}s can be found
\cite{Vilenkin1992b}.
\section{Clebsch-Gordan coefficients of lower-weight states}\label{sec:lower-weight}
Let us now turn to the \CGC{}s of states of $S''$ other than its highest-weight
state.  These are obtained by acting on both sides of Eq.~\eqref{eq:hwexp}
with lowering operators, using Eq.~\eqref{eq:jmelement} for the matrix
representations of $J^{(l)}_-$ for the carrier space $ \mathbb{V}^{S'',
\alpha}$ on the left-hand side, and for the direct product carrier space $
\mathbb{V}^S \otimes \mathbb{V}^{S'}$ on the right-hand side. However,
according to Eq.~\eqref{eq:jmelement}, the action of $J^{(l)}_-$ in general
produces not a unique basis state, but a linear combination of basis states of
$\mathbb{V}^{S'', \alpha}$. We shall therefore calculate, in parallel, the
\CGC{}s of \emph{all} basis states with a given $\alpha$ and given \pweight{} $W =
(w_l)$, i.e.\ of all $\ket{M'',\alpha}$ having $W(M'') = W$.

  To this end, assume that we have already determined all ``parent
  states'' of the desired \pweight{} $W$ within $\mathbb{V}^{S'',
    \alpha}$. By parent states we mean those which, when acted upon by
  a single $J^{(l)}_-$, yield (linear combinations of) states of
  weight $W$. For a given $J^{(l)}_-$ (with $1 \le l \le N-1$), the
  relevant parent states have \pweight{} $(w_1, \ldots, w_{l-1}, w_{l}+1,
  w_{l+1}-1, w_{l+2}, \ldots, w_N)$ and consist of all states of the
  form $\ket{M''+M^{k,l}, \alpha}$ with $W(M'') = W$ and $1 \le k \leq
  l$, for which $M'' + M^{k,l}$ is a valid \GTP{}.  Each parent state
  can be expressed as
\begin{equation}
  \ket{M'' + M^{k,l}, \alpha} = 
\sum_{M,M'} C^{M'' + M^{k,l}, \alpha}_{M,M'} \ket{M \otimes M'},
\label{eq:M''CGC}
\end{equation}
where the CGC{s} are, by assumption, already known. Now, the
action of $J^{(l)}_-$ on any parent state can be written as a linear
combination of all states $\ket{M''', \alpha}$ with $W(M''') = W$,
\begin{equation}
J^{(l),S''}_- \ket{M'' + M^{k,l}, \alpha} = \sum_{M'''} b^{M'''}_{M'',k,l} \ket{M''', \alpha},
\label{eq:JM''}
\end{equation}
where the coefficients $b^{M'''}_{M'',k,l}$ are determined by the
  matrix representation of $J^{(l)}_-$ within $ \mathbb{V}^{S'',
    \alpha}$, as given by Eq.~\eqref{eq:jmelement}.  Combining
  Eqs.~(\ref{eq:M''CGC}) and (\ref{eq:JM''}) and using the
direct product representation of $J^{(l)}_-$ on
$\mathbb{V}^S \otimes \mathbb{V}^{S'}$, we obtain a linear
  system of equations of the form
(compare Eq.~\eqref{eq:CGC-for-lower-weight}):
\begin{equation}
\label{eq:lower-system}
\sum_{M'''} b^{M'''}_{M'',k,l} \ket{M''', \alpha}
= \sum_{M,M'} C^{M'' + M^{k,l}, \alpha}_{M,M'}
(J^{(l),S}_- \otimes \mathbb{I}^{S'} + \mathbb{I}^{S} \otimes J^{(l),S'}_-)
\ket{M \otimes M'}.
\end{equation}
Each combination of indices $M''$, $k$, and $l$ specifies a separate equation,
where $M''$ runs over all \GTP{}s such that $W(M'') = W$, $l$ runs
from $1$ to $N-1$, and $k$ runs from $1$ to $l$, provided that $M'' +
M^{k,l}$ is a valid \GTP{}. Actually, only $I(W)$ of these equations
are linearly independent; as we do not know in advance which ones
these are, we include them all, i.e.\ the system of equations
(\ref{eq:lower-system}) is, in general, overdetermined.
Since the action of the $J^{(l)}_-$s on the right-hand side is known
from Eq.~\eqref{eq:jmelement},
the sought-after \CGC{}s $C^{M'',\alpha}_{M,M'}$ can now be readily obtained by
inverting the matrix of the coeffcients $b^{M'''}_{M'',k,l}$ in order
to bring Eq.~\eqref{eq:lower-system} into the familiar form of
Eq.~\eqref{eq:defineCGC}.
\section{Algorithm for computer implementation}\label{sec:algorithm}
Having gathered in the preceding sections all necessary ingredients,
we are now ready to formulate the
sought-after algorithm for calculating \SU N \CGC{}s.
Given two \SU N irreps $S$ and $S'$,
perform the following steps:
\begin{enumerate}
\item Find the irreps $S''$ appearing in the decomposition of $S \otimes S'$,
as described in Sec.~\ref{sec:decompose}.
\item For each irrep $S''$, find the Clebsch-Gordan coefficients of the
$N^{S''}_{S,S'}$ highest-weight states $\ket{H'',\alpha}$.
Resolve outer multiplicity ambiguities,
as described in Sec.~\ref{sec:outer-multiplicity}.
\item From each highest-weight state $\ket{H'',\alpha}$, construct the
lower-weight states by repeated application of $J^{(l)}_-$ operators, treating each weight of
$S''$ separately, as described in Sec.~\ref{sec:lower-weight}.
\end{enumerate}
An explicit computer implementation of this strategy
is presented in App.~\ref{app:source}.
To check that our algorithm works correctly,
we have verified that it satisfies the following consistency checks:
\begin{itemize}
\item For \SU 2 and \SU 3, the results coincide with known formulas and tables,
up to sign conventions.
\item The selection rule~\eqref{eq:sun-selection-rule} is fulfilled.
\item The matrix $C$ of Clebsch-Gordan coefficients (see Sec.~\ref{sec:defineCGC})
is unitary.
\item The matrix $C$ block-diagonalizes
the representation matrices (Eqs.~\eqref{eq:blockdiagonalize}).
\end{itemize}

The speed of the algorithm depends polynomially
on the dimensions of the irreps $S$ and $S'$.
On a modern computer (2~GHz CPU clock speed),
smaller \su 3 cases (e.g.\ $\dim S = 6, \dim S' = 15$) run instantly,
while medium-sized \su 5 cases (e.g. $\dim S = 35, \dim S' = 224$) take a few minutes,
and larger \su 5 cases (e.g. $\dim S = 280, \dim S' = 420$) require several hours computing time.

As an outlook, we note that it should be possible to greatly speed up
our algorithm by exploiting the fact that the weight diagrams are
symmetric under the Weyl group, which in this context can be thought
of as the group of all permutations of the elements of the \pweight{}s,
$(w_1^M, \ldots, w_N^M) \to (w_{\sigma(1)}^M, \ldots, w_{\sigma(N)}^M)$.
Exploiting this symmetry is a nontrivial
task, since the Gelfand-Tsetlin basis is not stable unter the
operation of the Weyl group. Nevertheless, we expect that it should be possible to
do within the general framework of our algorithm, by adopting a
suitably modified state labeling scheme that exploits the Weyl
symmetry.  Work along these lines is currently in progress.

\textbf{Acknowledgements:} This work was inspired by the progress made by
G.~Zar\'and, C.~P.~Moca and collaborators\cite{Toth2008} in devising a
flexible code for the numerical renormalization group, capable of exploiting
non-Abelian symmetries. We acknowledge stimulating and helpful discussions
with 
G.~Buchalla,
R.~Helling,
M.~Kieburg,
P.~Littelmann,
C.~P.~Moca,
A.~Weichselbaum,
and
G.~Zar\'and,
and financial support from SFB-TR12.
\appendix
\section{Correspondence between Gelfand-Tsetlin patterns and Young tableaux}
\label{app:young}
There exists a one-to-one correspondence between \iweight{}s and
  Young diagrams, and between \GTP{}s and semi-standard Young
  tableaux. Thus, our algorithm could equally well have been
  formulated in terms of Young diagrams and Young tableaux. Since the
  latter are easy to visualize and are perhaps more widely known in the
  physics community than the GT-scheme, this appendix summarizes the
  relation between the two schemes.  Our reason for preferring \GTP{}s 
 to Young tableaux lies in the complexity of the computer
  implementation: GT-patterns can be stored in a simpler data
  structure and allow for a simpler evaluation of the matrix
  elements~\eqref{eq:jmelement} and~\eqref{eq:jpelement}.

Note that Young tableaux can also be used to label bases that differ
from the GT basis used in this work, notably the one constructed via
Young symmetrizers \cite[ch.~7]{Lichtenberg1978}.  Thus, the
correspondence between \GTP{}s and Young tableaux set forth below is purely of
combinatorial nature.
\subsection{Definition of Young diagrams and Young tableaux}
\begin{table}
\mbox{\subref{tab:sy3yd}}
\subfloat{
  \label{tab:sy3yd}
  $\yng(1) \quad \yng(1,1) \quad \yng(2) \quad \yng(2,2) \quad \yng(4,2)
  \quad \yng(3,2,2)$
}
\hfill
\mbox{\subref{tab:sy3yt}}
\subfloat{
  \label{tab:sy3yt}
  $\young(11,2) \quad \young(11,3) \quad \young(12,2) \quad \young(12,3)
  \quad \young(13,2) \quad \young(13,3) \quad \young(22,3) \quad \young(23,3)$
}
\caption{
  \subref{tab:sy3yd} Examples of Young diagrams of \su 3 irreps. Since columns
  with 3 boxes can be deleted, the last example is effectively equal to the
  first one.
  \subref{tab:sy3yt} Set of all of valid \su 3 Young tableaux of
  shape~{\tiny$\protect\yng(2,1)$}.
}
\label{tab:young-examples}
\end{table}
A Young diagram is an arrangement of boxes in rows and columns
conforming to the following rules:
\begin{inparaenum}
\item[(YD.1)] there is a single, contiguous cluster of boxes;
\item[(YD.2)] the left borders of all rows are aligned; and
\item[(YD.3)] each row is not longer than the one above.
\end{inparaenum}

Note that the empty Young diagram consisting of no boxes is a valid
Young diagram. For the purpose of describing an \su N irrep,
we additionally require that
\begin{inparaenum}
\item[(YD.4)] there are at most $N$ rows; and
\item[(YD.5)] columns with $N$ boxes are dropped,
\end{inparaenum}
i.e.\ diagrams which differ only by such columns are identified with each other.

Every Young diagram $D$ satisfying rules (YD.1) to (YD.5) uniquely labels
an \su N irrep (or \slc N irrep), i.e.\ the label $S$ used in the main text
can be associated with a Young diagram $D$.
Some \su 3 examples are shown in Table~\ref{tab:sy3yd}.
A further example is given by the Young diagrams specifying \su 2 irreps:
The irrep $S = j$ (describing total angular momentum \(j\)) corresponds
to a Young diagram with \(2j\) boxes in a single row.

A \emph{(semi-standard) Young tableau} is a Young diagram,
of which the boxes are filled according to the following rules:
\begin{inparaenum}
\item[(YT.1)] \label{enum:yt1} Each box contains a single integer between 1 and \(N\), inclusive;
\item[(YT.2)] \label{enum:yt2} the numbers in each row of boxes weakly increase from left to right
(i.e.\ each number is equal to or larger than the one to its left); and
\item[(YT.3)] \label{enum:yt3} the numbers in each column strictly increase from top to bottom
(i.e. each number is strictly larger than the one above it).
\end{inparaenum}

The basis states of an \su N representation identified
by a given Young diagram $D$ can be uniquely labeled by the set of all 
associated valid semi-standard Young tableaux
(satisfying rules YT.1 to YT.3),
i.e.\ the label $M$ used in the main text
can be associated with a valid Young tableau $T$.
We shall denote the corresponding state by $\ket{T}$. 
For example, all eight Young tableaux for the diagram {\tiny$\yng(2,1)$}
with respect to \su 3 are shown in Table~\ref{tab:sy3yt}. As another example,
let us give the correspondence between states $\ket{S,m}$ of an \su 2 irrep
and Young tableaux: $\ket{S,m}$ corresponds to a Young tableau with $2S$ boxes
in a single row, containing $1$ in the leftmost $S+m$ boxes and $2$ in the
remaining $S-m$ boxes.

The dimension of a carrier space labeled by a Young diagram is given by the number of valid Young tableaux with the same shape as the Young diagram.
\subsection{Translating GT-patterns to Young tableaux} 
\label{sec:gt-young}
\begin{table}
\mbox{\subref{tab:rows}}
\subfloat{
  \label{tab:rows}
  \begin{tabular}{c|c|c|c}
    Row 1 & Row 2 & Row 3 & Row 4 \\
    \hline
    \vphantom{\fbox{\tiny$\begin{pmatrix}
      \multicolumn{2}{c}{4} & \multicolumn{2}{c}{3} & \multicolumn{2}{c}{1} & \multicolumn{2}{c}{0} \\
      & \multicolumn{2}{c}{3} & \multicolumn{2}{c}{2} & \multicolumn{2}{c}{1} & \\
      && \multicolumn{2}{c}{3} & \multicolumn{2}{c}{2} && \\
      &&& \multicolumn{2}{c}{2} &&&
    \end{pmatrix}$}}
    {\tiny$\begin{pmatrix}
      \multicolumn{2}{c}{\phantom{4}} & \multicolumn{2}{c}{\phantom{3}} & \multicolumn{2}{c}{\phantom{1}} & \multicolumn{2}{c}{\phantom{0}} \\
      & \multicolumn{2}{c}{\phantom{3}} & \multicolumn{2}{c}{\phantom{2}} & \multicolumn{2}{c}{\phantom{1}} & \\
      && \multicolumn{2}{c}{\phantom{3}} & \multicolumn{2}{c}{\phantom{2}} && \\
      &&& \multicolumn{2}{c}{2} &&&
    \end{pmatrix}$} & {\tiny$\begin{pmatrix}
      \multicolumn{2}{c}{\phantom{4}} & \multicolumn{2}{c}{\phantom{3}} & \multicolumn{2}{c}{\phantom{1}} & \multicolumn{2}{c}{\phantom{0}} \\
      & \multicolumn{2}{c}{\phantom{3}} & \multicolumn{2}{c}{\phantom{2}} & \multicolumn{2}{c}{\phantom{1}} & \\
      && \multicolumn{2}{c}{3} & \multicolumn{2}{c}{2} && \\
      &&& \multicolumn{2}{c}{2} &&&
    \end{pmatrix}$} & {\tiny$\begin{pmatrix}
      \multicolumn{2}{c}{\phantom{4}} & \multicolumn{2}{c}{\phantom{3}} & \multicolumn{2}{c}{\phantom{1}} & \multicolumn{2}{c}{\phantom{0}} \\
      & \multicolumn{2}{c}{3} & \multicolumn{2}{c}{2} & \multicolumn{2}{c}{1} & \\
      && \multicolumn{2}{c}{3} & \multicolumn{2}{c}{2} && \\
      &&& \multicolumn{2}{c}{2} &&&
    \end{pmatrix}$} & {\tiny$\begin{pmatrix}
      \multicolumn{2}{c}{4} & \multicolumn{2}{c}{3} & \multicolumn{2}{c}{1} & \multicolumn{2}{c}{0} \\
      & \multicolumn{2}{c}{3} & \multicolumn{2}{c}{2} & \multicolumn{2}{c}{1} & \\
      && \multicolumn{2}{c}{3} & \multicolumn{2}{c}{2} && \\
      &&& \multicolumn{2}{c}{2} &&&
    \end{pmatrix}$} \\
    \hline
    \vphantom{\fbox{$\yng(1,1,1)$}} $\young(11)$ & $\young(112,22)$ & $\young(112,22,3)$ & $\young(1124,224,3)$
  \end{tabular}
}
\hfill
\mbox{\subref{tab:diagonals}}
\subfloat{
  \label{tab:diagonals}
  \begin{tabular}{c|c|c|c}
    Diagonal 1 & Diagonal 2 & Diagonal 3 & Diagonal 4 \\
    \hline
    \vphantom{\fbox{\tiny$\begin{pmatrix}
      \multicolumn{2}{c}{4} & \multicolumn{2}{c}{3} & \multicolumn{2}{c}{1} & \multicolumn{2}{c}{0} \\
      & \multicolumn{2}{c}{3} & \multicolumn{2}{c}{2} & \multicolumn{2}{c}{1} & \\
      && \multicolumn{2}{c}{3} & \multicolumn{2}{c}{2} && \\
      &&& \multicolumn{2}{c}{2} &&&
    \end{pmatrix}$}}
    {\tiny$\begin{pmatrix}
      \multicolumn{2}{c}{4} & \multicolumn{2}{c}{\phantom{3}} & \multicolumn{2}{c}{\phantom{1}} & \multicolumn{2}{c}{\phantom{0}} \\
      & \multicolumn{2}{c}{3} & \multicolumn{2}{c}{\phantom{2}} & \multicolumn{2}{c}{\phantom{1}} & \\
      && \multicolumn{2}{c}{3} & \multicolumn{2}{c}{\phantom{2}} && \\
      &&& \multicolumn{2}{c}{2} &&&
    \end{pmatrix}$} & {\tiny$\begin{pmatrix}
      \multicolumn{2}{c}{4} & \multicolumn{2}{c}{3} & \multicolumn{2}{c}{\phantom{1}} & \multicolumn{2}{c}{\phantom{0}} \\
      & \multicolumn{2}{c}{3} & \multicolumn{2}{c}{2} & \multicolumn{2}{c}{\phantom{1}} & \\
      && \multicolumn{2}{c}{3} & \multicolumn{2}{c}{2} && \\
      &&& \multicolumn{2}{c}{2} &&&
    \end{pmatrix}$} & {\tiny$\begin{pmatrix}
      \multicolumn{2}{c}{4} & \multicolumn{2}{c}{3} & \multicolumn{2}{c}{1} & \multicolumn{2}{c}{\phantom{0}} \\
      & \multicolumn{2}{c}{{3}} & \multicolumn{2}{c}{{2}} & \multicolumn{2}{c}{{1}} & \\
      && \multicolumn{2}{c}{3} & \multicolumn{2}{c}{2} && \\
      &&& \multicolumn{2}{c}{2} &&&
    \end{pmatrix}$} & {\tiny$\begin{pmatrix}
      \multicolumn{2}{c}{4} & \multicolumn{2}{c}{3} & \multicolumn{2}{c}{1} & \multicolumn{2}{c}{0} \\
      & \multicolumn{2}{c}{3} & \multicolumn{2}{c}{2} & \multicolumn{2}{c}{1} & \\
      && \multicolumn{2}{c}{3} & \multicolumn{2}{c}{2} && \\
      &&& \multicolumn{2}{c}{2} &&&
    \end{pmatrix}$} \\
    \hline
    \vphantom{\fbox{$\yng(1,1,1)$}} $\young(1124)$ & $\young(1124,224)$ & $\young(1124,224,3)$ & $\young(1124,224,3)$
  \end{tabular}
}
\caption{Conversion of a \GTP{} to a Young tableau, stepping
  \subref{tab:rows} along rows, and
  \subref{tab:diagonals} along diagonals.
}
\label{tab:conversion}
\end{table}
Each \GTP{} $M = (m_{k,l})$ uniquely specifies a corresponding Young tableau
(p.~526 of Ref.~\onlinecite{Louck2008}), which
can be constructed as follows. Start with an empty Young tableau (no
boxes at all), and step through the entries of the pattern using
either of the following two stepping orders,
illustrated in Table~\ref{tab:rows} and~\ref{tab:diagonals},
respectively:
\begin{enumerate}[(a)]
\item Proceed from the bottom to top, one row at a time (increasing  $l$ from 1 to $n$), and within each row from left to right (increasing $k$ from 1 to $l$); or
\item Proceed from left to right, one diagonal at a time  (increasing $k$ from 1 to $n$), and within each diagonal from bottom to top (increasing $l$ from  $k$ to $n$).
\end{enumerate}
For each step to a new entry in the pattern, say $m_{k,l}$ located in
diagonal $k$ and row $l$, extend the length of the $k$-th tableau row
to a total of $m_{k,l}$ boxes, by adding to its right boxes containing
the number $l$.

According to the above procedure, the topmost row of the
  GT-pattern specifies the number of boxes in the rows of the
  corresponding Young diagram: for the latter, row $k$ of the latter contains
  $m_{k,N}$ boxes.  In this way, the information specifying the irrep
  $S$, which for a GT-pattern resides in its topmost row, specifies
  the shape of the corresponding Young diagram.
Moreover,  the number of $l$-boxes (i.e.\ boxes containing the number $l$) in tableau
row $k$, say $d_{k,l}$, is given by 
\begin{equation}
\label{eq:numberofl-boxesinrowk}
d_{k,l} = m_{k,l} - m_{k,l-1} \; , \qquad
(\text{where} \quad m_{k,l} \equiv 0  \; 
\text{if} \;  k > l ).
\end{equation}

Since both stepping orders ensure that pattern entries in the same
diagonal $k$ are visited in order of increasing $l$, they yield the
same final Young tableau. Order (b) has the feature that an entire
tableaux row is completed before the next row is begun. As a result,
(b) is more convenient for transcribing
the Littlewood-Richardson rule for decomposing a product representation
from the language of Young tableaux to that of \GTP{}s.

The converse process of transcribing a Young tableau to a \GTP{}
can be achieved by using the tableau's $k$-th row,
read from left to right, to fill in the pattern's $k$-th diagonal,
from bottom to top, in such way as to respect the above rules.
\subsection{Remarks about Young tableaux}
In order to aid our intuition for the \su N representation theory
presented in the main text, this section restates some of the
properties discussed there in terms of Young tableaux.

The \pweight{} $W(M)$ of a \GTP{} $M$, as introduced in Sec.~\ref{sec:weights},
has an illustrative interpretation; $w^M_l$ is the number of
$l$-boxes (i.e.\ boxes containing $l$)
in the tableau corresponding to $M$.
Thus, for the highest-weight Young tableau (the \GTP{} of the corresponding
state $\ket{H}$ is specified at the end of Sec.~\ref{sec:raising-lowering}),
row $l$ from the top contains only $l$-boxes (i.e.\ $w^M_l = m_{l,N}$),
e.g.\ {\tiny$\young(1111,22,33)$}.
Furthermore, if the states $\ket{T}$ and $\ket{T'}$ have the same \pweight{},
the tableaux $T$ and $T'$  contain the same set of entries
(i.e.\ the same number of $l$-boxes), but arranged in different ways.
For example, for \su 3 {\tiny$\young(12,3)$} and {\tiny$\young(13,2)$}
have the same \pweight{} $W = (1,1,1)$.

The action of the raising and lowering operators $J^{(l)}_\pm$ on \pweight{}s
is given by Eq.~\eqref{eq:weight-shift}.
The corresponding action of $J^{(l)}_+$ on a state labeled by a Young tableau $T$ produces
a linear combination of states labeled by tableaux containing one
more $l$-box and one less $(l+1)$-box.
Analogously, $J^{(l)}_-$ has the reverse effect on Young
tableaux: it produces a linear combination of tableaux containing
one less $l$-box and one more $(l+1)$-box.
\section{Derivation of our formulation of the Littlewood-Richardson rule}
\label{app:Littlewood-Richardson}
Our formulation of the Littlewood-Richardson rule in Section~\ref{sec:decompose}
is based on a version by van~Leeuwen\cite{Leeuwen2007},
formulated in terms of Young tableaux, which we outline here.
We then rephrase this in the language of Gelfand-Tsetlin patterns
to derive the method presented in Sec.~\ref{sec:decompose},
in particular Eq.~\eqref{eq:condition}.

Given two Young diagrams $D$ and $D'$,
write down all possible semistandard Young tableaux for $D$,
and for each such tableau (to be called the \emph{current tableau} below),
construct a corresponding Young diagram (to be called the \emph{trial
diagram} below) in the following manner:
\begin{enumerate}
\item Start the trial diagram as a fresh copy of $D'$.
\item Step through the boxes of the current tableau from right to left,
from top to bottom.
\item If the box encountered at a given step is an $l$-box,
add a box at the right end of row $l$ of the trial diagram.
\item If this produces a trial diagram that is no longer a valid
Young diagram (having a row longer than the one above), discard it
and start anew with the next tableau.
\item If, however, a valid Young diagram is constructed during each step,
the final Young diagram obtained after the last step represents an irrep
occurring in the decomposition of $D \otimes D'$.
\end{enumerate}
Let us now translate the above steps into the GT-scheme,
thus deriving the rules set forth in Section~\ref{sec:decompose}. There,
we assume two \iweight{}s $S$ and $S'$
to be given instead of two Young diagrams.
Naturally, taking a fresh copy of $D'$ corresponds to initializing
$(t_1, \ldots, t_N) = (m'_{1,N}, \ldots, m'_{N,N})$,
and stepping through the current tableau in the reading order of step~2
corresponds to stepping through the \GTP{} $M$ associated with $S$
along the diagonals from top to bottom and from left to right.
(This follows from the rules for translating \GTP{}s to Young tableaux
given in Sec.~\ref{sec:gt-young}; recall that the $k$-th diagonal of a \GTP{}
specifies the content of the $k$-th row of the corresponding Young tableau.)

Instead of processing one box of the current tableau at a time,
we treat all identical boxes of a given row at once
when stepping through the corresponding \GTP{}.
Recalling that $b_{k,l}$ 
of Eq.~\eqref{eq:numberofl-boxesinrowk}
gives the number of $l$-boxes in row
$k$ of the current tableau, it follows that
$B_{k,l} \equiv m'_{l,N} + \sum_{k'=1}^{k} b_k$ then
gives the number of boxes in row $l$ of the trial diagram
after having processed all boxes of type $l$ in row $k$
of the current tableau.

The condition~\eqref{eq:condition}, which must be fulfilled
for all $1 \le k \le l \le N$, finally assures that the trial
diagram is a valid Young diagram after each step.
\section{Identifying irreps and states by a single integer}
\label{app:map}
For numerical codes dealing with \iweight{}s, it is useful to identify
each \iweight{} by a unique number. To this end, we need a one-to-one
mapping between the set of all \su N \iweight{}s (for given $N$) and the
set of nonnegative integers. We shall construct such a mapping by
devising an ordering rule for \iweight{}s, using this rule to arrange all
possible diagrams in a list of increasing order, and labeling each
\iweight{} by its position in this list.

Similarly, we would like to map \GTP{}s to matrix indices, so we also
need a one-to-one mapping between the set of all \GTP{}s belonging to
a given irrep and the integers from 1 to the dimension of that
irrep. Therefore, we also define an order on \GTP{}s of a given irrep
and proceed analogously.
\subsection{Identifying \iweight{}s with a single number}
\begin{table}
  \begin{center}
    \begin{minipage}{0.3\linewidth}
      \begin{tabular}{c|c}
        $P(S)$ & $S = (m_{1,4}, m_{2,4}, m_{3,4}, m_{4,4})$ \\
        \hline 
        0 & $(0, 0, 0, 0)$ \\
        1 & $(1, 0, 0, 0)$ \\
        2 & $(1, 1, 0, 0)$ \\
        3 & $(1, 1, 1, 0)$ \\
        4 & $(2, 0, 0, 0)$ \\
      \end{tabular}
    \end{minipage}
    \qquad
    \begin{minipage}{0.3\linewidth}
      \begin{tabular}{c|c}
        $P(S)$ & $S = (m_{1,4}, m_{2,4}, m_{3,4}, m_{4,4})$ \\
        \hline 
        5 & $(2, 1, 0, 0)$ \\
        6 & $(2, 1, 1, 0)$ \\
        7 & $(2, 2, 0, 0)$ \\
        8 & $(2, 2, 1, 0)$ \\
        9 & $(2, 2, 2, 0)$
      \end{tabular}
    \end{minipage}
  \end{center}
  \caption{The first few \iweight{}s of \su 4 (excluding
    weights with $m_{4,4} \ne 0$), arranged in increasing order.}
  \label{tab:first-few-su4}
\end{table}
\begin{figure}
    \subfloat{
      \label{fig:countingPk}
      \mbox{\subref{fig:countingPk}}
      \parbox[c]{0.4\linewidth}{\includegraphics[width=\linewidth]{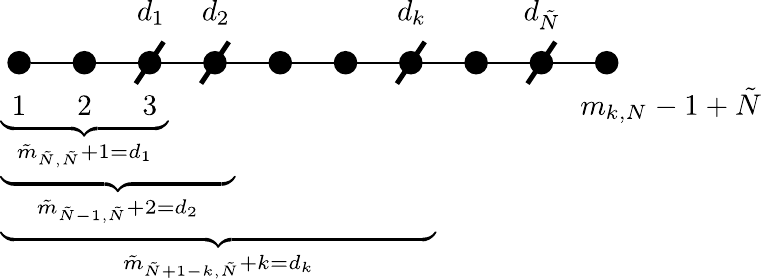}}
    }
    \quad
    \subfloat{
      \label{fig:countingPk2}
      \mbox{\subref{fig:countingPk2}}
      \parbox[c]{0.4\linewidth}{\includegraphics[width=\linewidth]{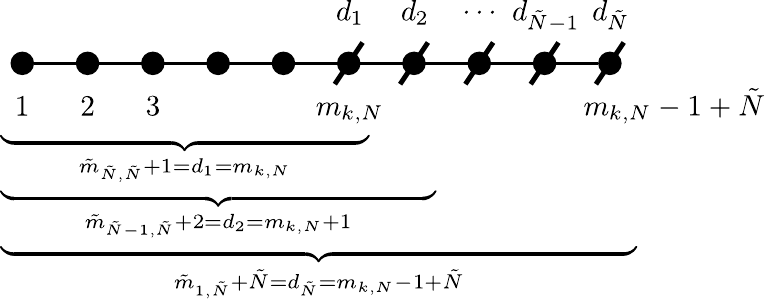}}
    }
  \caption{
    Enumeration scheme of \iweight{}s.
    \subref{fig:countingPk} Illustration of the combinatorics underlying $P_k(S)$.
    \subref{fig:countingPk2} Striking out items such that each
    $\tilde m_{\tilde k,\tilde N}$ takes on the largest possible value.
  }
  \label{fig:young-enumeration}
\end{figure}
We adopt throughout the convention for an \iweight{} $S = (m_{k,N})$ that
$m_{N,N} = 0$ (Sec.~\ref{sec:gelfand}).

For \iweight{}s we choose the following ordering rule: the ``smaller'' of
two \iweight{}s is taken to be the one with the smaller first element; in
case of a tie, compare the second element, and so on. Formally, given
two \iweight{}s $S$ and $S'$, we assign the 
order  
\begin{equation}
\label{eq:orderingS}
S < S' \; {\rm if \, and \, only \, if, \,
for \, the \, smallest \, index \, (say \,} k{\rm )\, for \, which \;}
m_{k,N} \neq m'_{k,N}, \; {\rm we \, have \;} m_{k,N} < m'_{k,N} \; . 
\end{equation}
Table~\ref{tab:first-few-su4} shows the
first few \iweight{}s of \SU 4, arranged in increasing order.

Using this ordering rule, all possible \su N \iweight{}s can be arranged
in a list of increasing order and uniquely labeled by a nonnegative
integer, say $P(S)$, giving its position in this list,
\begin{equation}
P(S) = \#\{S' | S' < S\}.
\end{equation}
To determine $P(S)$ for a given \iweight{} $S$, we simply count the number of smaller
weights $S'$: this number is given by the number (say $P_1(S)$) of
all weights $S'$ with $m'_{1,N} < m_{1,N}$, plus the number of all
$S'$ with $m'_{1,N} = m_{1,N}$ but $m'_{2,N} < m_{2,N}$ (say
$P_2(S)$), etc. Thus,
\begin{equation}
P(S) = \sum_{k = 1}^{N-1} P_k(S) \; ,
\end{equation}
where $P_k(S)$ is the number of weights $S'$ whose first $k-1$
entries are the same as those of $S$ ($m'_{k',N} = m_{k',N}$ for all
$k' < k$), while the $k$-th entry is arbitrary but smaller
than that of $S$ ($m'_{k,N} < m_{k,N}$), and the remaining entries
arbitrary (but subject to $S'$ being a valid \iweight{}, with
  $m'_{N,N} = 0$). The nontrivial ''free'' (though constrained) entries of $S'$,
 namely  $(m'_{k,N}, m'_{k+1,N}, \dots, m'_{N-1,N})$, can be viewed as an
\iweight{} $\tilde S = (\tilde m_{\tilde k, \tilde N}) $ of length
  $\tilde N = N-k$, whose entries $\tilde m_{\tilde k, \tilde N} =
  m'_{k-1+\tilde k, N}$ (for $1 \le \tilde k \le \tilde N$) satisfy
\begin{equation}
\label{eq:rules-for-tildem}
m_{k,N} -1  \ge \tilde m_{1,\tilde N} \ge \tilde m_{2, \tilde N} \ge \dots \ge \tilde m_{\tilde N, \tilde N} \ge  0 \; . 
\end{equation}
$P_k(S)$ thus is the number of allowed weights
$\tilde S$ that satisfy (\ref{eq:rules-for-tildem}).

To calculate $P_k(S)$, we note that it is equal to the number of ways
to draw or ``strike out'', from the set of integers $\{1,  \dots,
m_{k,N} -1 + \tilde N \}$, an ordered subset $\{d_{\tilde k} \}$ of
$\tilde N$ integers, $d_1, < d_2 < \dots < d_{\tilde N}$ (see
Fig.~\ref{fig:countingPk}), since there is a one-to-one correspondence
between the set of all possible such strike-outs and the set of all
\iweight{}s $\tilde S$ satisfying (\ref{eq:rules-for-tildem}): for a
given struck-out set $\{d_{\tilde k}\}$, with $1 \le \tilde k \le
\tilde N$, set $\tilde m_{\tilde k, \tilde N}$ equal to the number of
non-struck-out integers smaller than $d_{\tilde N+1-\tilde k}$
(i.e.\ $\tilde m_{\tilde k, \tilde N} = d_{\tilde N+1-\tilde k} -
(\tilde N+1-\tilde k)$).  For example, the weight $\tilde S$ that is
largest (w.r.t.\ to the ordering rule (\ref{eq:orderingS})), namely
having all elements equal to $m_{k,N} - 1$, is
obtained by choosing the struck-out integers $d_{\tilde k}$ to be as
large as possible (see Fig.~\ref{fig:countingPk2}). Thus, we have
\begin{equation}
P_k(S) = \binom{ m_{k,N} - 1 + \tilde N}{\tilde N} = \binom{N - k + m_{k,N} - 1}{N - k} \;,
\end{equation}
and, consequently,
\begin{equation}
P(S) = \sum_{k=1}^{N-1} \binom{N - k + m_{k,N} - 1}{N - k} \;.
\end{equation}
\subsection{Mapping of Gelfand-Tsetlin patterns to matrix indices}
\label{app:yt-mapping}
In analogy to the ordering we have defined on \iweight{}s, we introduce an
ordering on the set of Gelfand-Tsetlin patterns of a given irrep
(i.e.\ given top row of the pattern).  Let $M = (m_{k,l})$ and $M' =
(m'_{k,l})$ (where $1 \leq k \leq l \leq N$) denote two patterns with
$m_{k,N} = m'_{k,N}$ for $k = 1, \ldots, N$.  We define a
  row-by-row ordering of indices (see Table~\ref{tab:index-order}),
  increasing from left to right within a row, and from top row to
  bottom row, i.e.\ $(k,l) < (k',l')$ if $l = l'$ and $k < k'$, or if
  $l > l'$.  We then define $M' < M$ if and only if for the smallest
index for which $m'_{k,l} \ne m_{k,l}$, we have $m'_{k,l} < m_{k,l}$.
An example of this ordering is given in Table~\ref{tab:gt-ordering}.
\begin{table}
\subfloat{
    \label{tab:index-order}
    \mbox{\subref{tab:index-order}}
    \tiny
    \gt{1}{2}{3}{4}{5}{6}
}
\hfill
\subfloat{
    \label{tab:gt-ordering}
    \mbox{\subref{tab:gt-ordering}}
    \tiny
    \begin{tabular}{rccccccccccccccc}
      &   \gt{2}{1}{0}{1}{0}{0} & < & \gt{2}{1}{0}{1}{0}{1} & < & \gt{2}{1}{0}{1}{1}{1} 
      & < & \gt{2}{1}{0}{2}{0}{0} & < & \gt{2}{1}{0}{2}{0}{1} & < & \gt{2}{1}{0}{2}{0}{2}
      & < & \gt{2}{1}{0}{2}{1}{1} & < & \gt{2}{1}{0}{2}{1}{2} \\
      \rule{0pt}{1.5em}$Q(M):$ & 1 && 2 && 3 && 4 && 5 && 6 && 7 && 8
    \end{tabular}
}
\caption{\subref{tab:index-order} Illustrating the row-by-row rule chosen in
  App.~\ref{app:yt-mapping} to define an ordering scheme for the indices of
  \GTP{}s: $(k,l) < (k', l')$ if $l = l'$  and $k < k'$, or if $l> l'$.
  \subref{tab:gt-ordering} Ordering of all \GTP{}s belonging to the \SU 3
  irrep $(2,1,0)$, together with the corresponding pattern indices $Q(M)$.}
\end{table}

We map each Gelfand-Tsetlin pattern $M$ to a nonnegative integer $Q(M)$
by counting the number of smaller Gelfand-Tsetlin patterns, i.e.
\begin{equation}
Q(M) = \#\{M' | M' \le M\} \;.
\end{equation}
This number can be determined by generating the pattern (say $\tilde
M(\{\tilde m_{k,l}\})$) located directly preceding $M$ in the
  ordered list of patterns, then the pattern preceding $\tilde M$,
and so on, until we arrive at the beginning of this list.  To
construct the predecessor of the pattern $M$, we start by finding the
largest index $(\tilde k,\tilde l)$ whose entry $m_{\tilde k, \tilde l}$ can be
  decreased without violating the betweenness condition
  \eqref{eq:betweenness},  
rewritten here as 
\begin{align}
\label{eq:betweennessnew}
m_{ k, l+1} &\geq m_{k, l} 
\geq m_{ k+1, l + 1} & (1 \leq k < l+1 \leq N),
\end{align}
with respect to  smaller indices while disregarding it with respect
to larger indices (i.e.\ without violating the second inequality,
but disregarding the first).
Thus, $(\tilde k, \tilde l)$ is the index for which  $m_{k,l} =
  m_{k+1,l+1}$ for all $(k,l) > (\tilde k,\tilde l)$ but
  $m_{\tilde k,\tilde l} > m_{\tilde k+1,\tilde l+1}$.
 We then decrease $m_{\tilde k, \tilde l}$ by one and reset the
entries of all larger indices to the maximal values that satisfy the
new betweenness condition. Concretely:
\begin{equation}
  \tilde m_{k,l} = \begin{cases}
    m_{k,l} & \text{for} \; (k,l) < (\tilde k,\tilde l) \quad \text{(keep entries with smaller indices unchanged)} \\
    m_{k,l}-1 & \text{for} \; (k,l) = (\tilde k,\tilde l) \quad \text{(decrease by 1 the entry with largest index for which this is possible)} \\
    \tilde m_{k,l+1} & \text{for} \; (k,l) > (\tilde k,\tilde l) \quad \text{(give entries with larger indices their largest possible value).}
  \end{cases}
\end{equation}
The number $Q(M)$ is, of course, the number of times we can repeat the
process of constructing a preceding pattern.  This procedure maps the
lowest-weight and highest-weight states of an irrep $S$ to the numbers
1 and  dim$(S)$, respectively.
\section{Source code}
\label{app:source}
Below, we provide a C++ implementation of our algorithm,
consisting of four fundamental classes:
\begin{inparaenum}
\item  \texttt{weight} is a data structure for irrep and pattern weights,
\item \texttt{pattern} stores \GT{} patterns,
\item \texttt{decomposition} implements the Littlewood-Richardson rule, and
\item \texttt{coefficients} computes and stores the actual \CGC{}s.
\end{inparaenum}
The end of the source code contains examples of typical applications.
For example, to calculate \CGC{}s, perfom the following steps:
\begin{enumerate}
\item Create two objects  \texttt{clebsch::weight S} and 
 \texttt{clebsch::weight Sprime},  representing the irreps $S$ and $S'$.
\item Create the object  \texttt{decomp} as 
\texttt{clebsch::decomposition decomp(S, Sprime)}; this generates
the irreps $S''$ that occur in the  decomposition of $S
  \otimes S'$ according to the Littlewood-Richardson rule.
(Its output can be read out, if desired, as follows:
Read out the total number of irreps $S''$ by calling
\texttt{decomp.size()}. Read out the $j$-th one of these
(with $1 \le j \le$ \texttt{decomp.size()}) by
creating an object \texttt{clebsch::weight Sdoubleprime(decomp(j))}.
Read out its outer multiplicity by calling \texttt{decomp.multiplicity(Sdoubleprime)}.)
\item Pick one of these irreps \texttt{Sdoubleprime} and create the
  object \texttt{C} as \texttt{clebsch::coefficients C(Sdoubleprime, S, Sprime)}; this 
generates all CGCs $C^{M'',\alpha}_{MM'}$ needed for constructing  the irrep $S''$, with
multiplicity index $\alpha$, 
from  $S$ and $S'$.
\item The Clebsch-Gordan coefficient $C^{M'',\alpha}_{MM'}$ is then 
read out as \texttt{C(alpha, Qdoubleprime, Q, Qprime)},
where \texttt{alpha} indexes the outer multiplicity of $S''$, 
and \texttt{Q}, \texttt{Qprime}, and \texttt{Qdoubleprime}
are the pattern indices of $M$, $M'$ and $M''$. 
\end{enumerate}
Other common applications involve the translation between an
\iweight{} $S$ and its index $P(S)$, or between a \GTP{} $M$ and its
index $Q(M)$.  To obtain the i-weight index $P(S)$ from the object
\texttt{clebsch::weight S}, call \texttt{S.index()}, and to obtain the
pattern index $Q(M)$ from the object \texttt{clebsch::pattern M}, call
\texttt{M.index()}. Conversely, to construct an \iweight{} $S=(m_{k,N})$
from a given irrep index $P$, create the object
\texttt{clebsch::weight S(N,P)}, and read out the elements $m_{k,N}$
as \texttt{S(k)}.  Similarly, to construct a pattern $M = (m_{k,l})$
in irrep $S$ from a given pattern index $Q$, create the object
\texttt{clebsch::pattern M(S,Q)}, and read out the elements $m_{k,l}$
as \texttt{M(k,l)}. Finally, to find the dimension $d_S$ of the irrep $S$,
create the object \texttt{clebsch::weight S} and call \texttt{S.dimension()}. 

All of these applications are elaborated in the sample routine $\texttt{main}$ at the end
of the source code (starting around line 1000). They are also implemented in
the interactive CGC-generator available at \url{http://homepages.physik.uni-muenchen.de/~vondelft/Papers/ClebschGordan/}.

To locate the implementation of key equations of the algorithm
in the source code, search for the following equation numbers:
\begin{inparaitem}[]
\item \iweight{}s $S$: Eq.~(19);
\item GT-patterns $M$: Eq.~(20);
\item irrep dimension $\dim(S)$: Eq.~(22);
\item \pweight{}s $W(M)$: Eq.~(25);
\item raising and lowering operators $J^{(l)}_{\pm}$: Eqs.~(28) and~(29);
\item Littlewood-Richardson rule: Eq.~(31);
\item highest-weight \CGC{}s $C^{H'',\alpha}_{M,M'}$: Eq.~(36);
\item normal form of highest-weight \CGC{}s: Eq.~(37);
\item lower-weight \CGC{}s $C^{M'',\alpha}_{M,M'}$: Eq.~(40);
\item irrep index $P(S)$: Eq.~(C2);
\item pattern index $Q(M)$: Eq.~(C7).
\end{inparaitem}

To compile, type \texttt{g++ clebsch.cpp -llapack -lblas} on Linux,
or \texttt{g++ clebsch.cpp -framework vecLib} on Mac~OS~X.
On other operating systems, make sure that LAPACK is included in
the linking process. To achieve that, you may have to modify
the declaration of the funtions \texttt{dgesvd} and \texttt{dgels}.

\begin{lstlisting}
#include <algorithm>
#include <cassert>
#include <cmath>
#include <cstdio>
#include <cstdlib>
#include <cstring>
#include <functional>
#include <fstream>
#include <iostream>
#include <map>
#include <numeric>
#include <vector>

// Declaration of LAPACK subroutines
// Make sure the data types match your version of LAPACK

extern "C" void dgesvd_(char const* JOBU,
                        char const* JOBVT,
                        int const* M,
                        int const* N,
                        double* A,
                        int const* LDA,
                        double* S,
                        double* U,
                        int const* LDU,
                        double* VT,
                        int const* LDVT,
                        double* WORK,
                        int const* LWORK,
                        int *INFO);

extern "C" void dgels_(char const* TRANS,
                       int const* M,
                       int const* N,
                       int const* NRHS,
                       double* A,
                       int const* LDA,
                       double* B,
                       int const* LDB,
                       double* WORK,
                       int const* LWORK,
                       int *INFO);

namespace clebsch {
    const double EPS = 1e-12;

    // binomial coefficients
    class binomial_t {
        std::vector<int> cache;
        int N;

    public:
        int operator()(int n, int k);
    } binomial;

    // Eq. (19) and (25)
    class weight {
        std::vector<int> elem;

    public:
        // the N in "SU(N)"
        const int N;

        // create a non-initialized weight
        weight(int N);

        // create irrep weight of given index
        // Eq. (C2)
        weight(int N, int index);

        // assign from another instance
        clebsch::weight &operator=(const clebsch::weight &w);

        // access elements of this weight (k = 1, ..., N)
        int &operator()(int k);
        const int &operator()(int k) const;

        // compare weights
        // Eq. (C1)
        bool operator<(const weight &w) const;
        bool operator==(const weight &w) const;

        // element-wise sum of weights
        clebsch::weight operator+(const weight &w) const;

        // returns the index of this irrep weight (index = 0, 1, ...)
        // Eq. (C2)
        int index() const;

        // returns the dimension of this irrep weight
        // Eq. (22)
        long long dimension() const;
    };

    // Eq. (20)
    class pattern {
        std::vector<int> elem;

    public:
        // the N in "SU(N)"
        const int N;

        // copy constructor
        pattern(const pattern &pat);

        // create pattern of given index from irrep weight
        // Eq. (C7)
        pattern(const weight &irrep, int index = 0);

        // access elements of this pattern (l = 1, ..., N; k = 1, ..., l)
        int &operator()(int k, int l);
        const int &operator()(int k, int l) const;

        // find succeeding/preceding pattern, return false if not possible
        // Eq. (C9)
        bool operator++();
        bool operator--();

        // returns the pattern index (index = 0, ..., dimension - 1)
        // Eq. (C7)
        int index() const;

        // returns the pattern weight
        // Eq. (25)
        clebsch::weight get_weight() const;

        // returns matrix element of lowering operator J^(l)_-
        // between this pattern minus M^(k,l) and this pattern
        // (l = 1, ..., N; k = 1, ..., l)
        // Eq. (28)
        double lowering_coeff(int k, int l) const;

        // returns matrix element of raising operator J^(l)_+
        // between this pattern plus M^(k,l) and this pattern
        // (l = 1, ..., N; k = 1, ..., l)
        // Eq. (29)
        double raising_coeff(int k, int l) const;
    };

    class decomposition {
        std::vector<clebsch::weight> weights;
        std::vector<int> multiplicities;

    public:
        // the N in "SU(N)"
        const int N;

        // save given irreps for later use
        const weight factor1, factor2;

        // construct the decomposition of factor1 times factor2 into irreps
        // Eq. (31)
        decomposition(const weight &factor1, const weight &factor2);

        // return the number of occurring irreps
        int size() const;

        // access the occurring irreps
        // j = 0, ..., size() - 1
        const clebsch::weight &operator()(int j) const;

        // return the outer multiplicity of irrep in this decomposition
        int multiplicity(const weight &irrep) const;
    };

    class index_adapter {
        std::vector<int> indices;
        std::vector<int> multiplicities;

    public:
        // the N in "SU(N)"
        const int N;

        // save given irreps for later use
        const int factor1, factor2;

        // construct this index_adapter from a given decomposition
        index_adapter(const clebsch::decomposition &decomp);

        // return the number of occurring irreps
        int size() const;

        // access the occurring irreps
        int operator()(int j) const;

        // return the outer multiplicity of irrep in this decomposition
        int multiplicity(int irrep) const;
    };

    class coefficients {
        std::map<std::vector<int>, double> clzx;

        // access Clebsch-Gordan coefficients in convenient manner
        void set(int factor1_state,
                 int factor2_state,
                 int multiplicity_index,
                 int irrep_state,
                 double value);

        // internal functions, doing most of the work
        void highest_weight_normal_form(); // Eq. (37)
        void compute_highest_weight_coeffs(); // Eq. (36)
        void compute_lower_weight_coeffs(int multip_index, int state, std::vector<char> &done); // Eq. (40)

    public:
        // the N in "SU(N)"
        const int N;

        // save irreps and their dimensions for later use
        const weight factor1, factor2, irrep;
        const int factor1_dimension, factor2_dimension, irrep_dimension;

        // outer multiplicity of irrep in this decomposition
        const int multiplicity;

        // construct all Clebsch-Gordan coefficients of this decomposition
        coefficients(const weight &irrep, const weight &factor1, const weight &factor2);

        // access Clebsch-Gordan coefficients (read-only)
        // multiplicity_index = 0, ..., multiplicity - 1
        // factor1_state = 0, ..., factor1_dimension - 1
        // factor2_state = 0, ..., factor2_dimension - 1
        // irrep_state = 0, ..., irrep_dimension
        double operator()(int factor1_state,
                          int factor2_state,
                          int multiplicity_index,
                          int irrep_state) const;
    };
};

// implementation of "binomial_t" starts here

int clebsch::binomial_t::operator()(int n, int k) {
    if (N <= n) {
        for (cache.resize((n + 1) * (n + 2) / 2); N <= n; ++N) {
            cache[N * (N + 1) / 2] = cache[N * (N + 1) / 2 + N] = 1;
            for (int k = 1; k < N; ++k) {
                cache[N * (N + 1) / 2 + k] = cache[(N - 1) * N / 2 + k]
                                           + cache[(N - 1) * N / 2 + k - 1];
            }
        }
    }

    return cache[n * (n + 1) / 2 + k];
}

// implementation of "weight" starts here

clebsch::weight::weight(int N) : elem(N), N(N) {}

clebsch::weight::weight(int N, int index) : elem(N, 0), N(N) {
    for (int i = 0; index > 0 && i < N; ++i) {
        for (int j = 1; binomial(N - i - 1 + j, N - i - 1) <= index; j <<= 1) {
            elem[i] = j;
        }

        for (int j = elem[i] >> 1; j > 0; j >>= 1) {
            if (binomial(N - i - 1 + (elem[i] | j), N - i - 1) <= index) {
                elem[i] |= j;
            }
        }

        index -= binomial(N - i - 1 + elem[i]++, N - i - 1);
    }
}

clebsch::weight &clebsch::weight::operator=(const clebsch::weight &w) {
    int &n = const_cast<int &>(N);
    elem = w.elem;
    n = w.N;
    return *this;
}

int &clebsch::weight::operator()(int k) {
    assert(1 <= k && k <= N);
    return elem[k - 1];
}

const int &clebsch::weight::operator()(int k) const {
    assert(1 <= k && k <= N);
    return elem[k - 1];
}

bool clebsch::weight::operator<(const weight &w) const {
    assert(w.N == N);
    for (int i = 0; i < N; ++i) {
        if (elem[i] - elem[N - 1] != w.elem[i] - w.elem[N - 1]) {
            return elem[i] - elem[N - 1] < w.elem[i] - w.elem[N - 1];
        }
    }
    return false;
}

bool clebsch::weight::operator==(const weight &w) const {
    assert(w.N == N);

    for (int i = 1; i < N; ++i) {
        if (w.elem[i] - w.elem[i - 1] != elem[i] - elem[i - 1]) {
            return false;
        }
    }

    return true;
}

clebsch::weight clebsch::weight::operator+(const weight &w) const {
    weight result(N);

    transform(elem.begin(), elem.end(), w.elem.begin(), result.elem.begin(), std::plus<int>());

    return result;
}

int clebsch::weight::index() const {
    int result = 0;

    for (int i = 0; elem[i] > elem[N - 1]; ++i) {
        result += binomial(N - i - 1 + elem[i] - elem[N - 1] - 1, N - i - 1);
    }

    return result;
}

long long clebsch::weight::dimension() const {
    long long numerator = 1, denominator = 1;

    for (int i = 1; i < N; ++i) {
        for (int j = 0; i + j < N; ++j) {
            numerator *= elem[j] - elem[i + j] + i;
            denominator *= i;
        }
    }

    return numerator / denominator;
}

// implementation of "pattern" starts here

clebsch::pattern::pattern(const pattern &p) : elem(p.elem), N(p.N) {}

clebsch::pattern::pattern(const weight &irrep, int index) :
        elem((irrep.N * (irrep.N + 1)) / 2), N(irrep.N) {
    for (int i = 1; i <= N; ++i) {
        (*this)(i, N) = irrep(i);
    }

    for (int l = N - 1; l >= 1; --l) {
        for (int k = 1; k <= l; ++k) {
            (*this)(k, l) = (*this)(k + 1, l + 1);
        }
    }

    while (index-- > 0) {
        bool b = ++(*this);

        assert(b);
    }
}

int &clebsch::pattern::operator()(int k, int l) {
    return elem[(N * (N + 1) - l * (l + 1)) / 2 + k - 1];
}

const int &clebsch::pattern::operator()(int k, int l) const {
    return elem[(N * (N + 1) - l * (l + 1)) / 2 + k - 1];
}

bool clebsch::pattern::operator++() {
    int k = 1, l = 1;

    while (l < N && (*this)(k, l) == (*this)(k, l + 1)) {
        if (--k == 0) {
            k = ++l;
        }
    }

    if (l == N) {
        return false;
    }

    ++(*this)(k, l);

    while (k != 1 || l != 1) {
        if (++k > l) {
            k = 1;
            --l;
        }

        (*this)(k, l) = (*this)(k + 1, l + 1);
    }

    return true;
}

bool clebsch::pattern::operator--() {
    int k = 1, l = 1;

    while (l < N && (*this)(k, l) == (*this)(k + 1, l + 1)) {
        if (--k == 0) {
            k = ++l;
        }
    }

    if (l == N) {
        return false;
    }

    --(*this)(k, l);

    while (k != 1 || l != 1) {
        if (++k > l) {
            k = 1;
            --l;
        }

        (*this)(k, l) = (*this)(k, l + 1);
    }

    return true;
}

int clebsch::pattern::index() const {
    int result = 0;

    for (pattern p(*this); --p; ++result) {}

    return result;
}

clebsch::weight clebsch::pattern::get_weight() const {
    clebsch::weight result(N);

    for (int prev = 0, l = 1; l <= N; ++l) {
        int now = 0;

        for (int k = 1; k <= l; ++k) {
            now += (*this)(k, l);
        }

        result(l) = now - prev;
        prev = now;
    }

    return result;
}

double clebsch::pattern::lowering_coeff(int k, int l) const {
    double result = 1.0;

    for (int i = 1; i <= l + 1; ++i) {
        result *= (*this)(i, l + 1) - (*this)(k, l) + k - i + 1;
    }
    
    for (int i = 1; i <= l - 1; ++i) {
        result *= (*this)(i, l - 1) - (*this)(k, l) + k - i;
    }

    for (int i = 1; i <= l; ++i) {
        if (i == k) continue;
        result /= (*this)(i, l) - (*this)(k, l) + k - i + 1;
        result /= (*this)(i, l) - (*this)(k, l) + k - i;
    }

    return std::sqrt(-result);
}

double clebsch::pattern::raising_coeff(int k, int l) const {
    double result = 1.0;

    for (int i = 1; i <= l + 1; ++i) {
        result *= (*this)(i, l + 1) - (*this)(k, l) + k - i;
    }

    for (int i = 1; i <= l - 1; ++i) {
        result *= (*this)(i, l - 1) - (*this)(k, l) + k - i - 1;
    }

    for (int i = 1; i <= l; ++i) {
        if (i == k) continue;
        result /= (*this)(i, l) - (*this)(k, l) + k - i;
        result /= (*this)(i, l) - (*this)(k, l) + k - i  - 1;
    }

    return std::sqrt(-result);
}

// implementation of "decomposition" starts here

clebsch::decomposition::decomposition(const weight &factor1, const weight &factor2) :
        N(factor1.N), factor1(factor1), factor2(factor2) {
    assert(factor1.N == factor2.N);
    std::vector<clebsch::weight> result;
    pattern low(factor1), high(factor1);
    weight trial(factor2);
    int k = 1, l = N;

    do {
        while (k <= N) {
            --l;
            if (k <= l) {
                low(k, l) = std::max(high(k + N - l, N), high(k, l + 1) + trial(l + 1) - trial(l));
                high(k, l) = high(k, l + 1);
                if (k > 1 && high(k, l) > high(k - 1, l - 1)) {
                    high(k, l) = high(k - 1, l - 1);
                }
                if (l > 1 && k == l && high(k, l) > trial(l - 1) - trial(l)) {
                    high(k, l) = trial(l - 1) - trial(l);
                }
                if (low(k, l) > high(k, l)) {
                    break;
                }
                trial(l + 1) += high(k, l + 1) - high(k, l);
            } else {
                trial(l + 1) += high(k, l + 1);
                ++k;
                l = N;
            }
        }

        if (k > N) {
            result.push_back(trial);
            for (int i = 1; i <= N; ++i) {
                result.back()(i) -= result.back()(N);
            }
        } else {
            ++l;
        }

        while (k != 1 || l != N) {
            if (l == N) {
                l = --k - 1;
                trial(l + 1) -= high(k, l + 1);
            } else if (low(k, l) < high(k, l)) {
                --high(k, l);
                ++trial(l + 1);
                break;
            } else {
                trial(l + 1) -= high(k, l + 1) - high(k, l);
            }
            ++l;
        }
    } while (k != 1 || l != N);

    sort(result.begin(), result.end());
    for (std::vector<clebsch::weight>::iterator it = result.begin(); it != result.end(); ++it) {
        if (it != result.begin() && *it == weights.back()) {
            ++multiplicities.back();
        } else {
            weights.push_back(*it);
            multiplicities.push_back(1);
        }
    }
}

int clebsch::decomposition::size() const {
    return weights.size();
}

const clebsch::weight &clebsch::decomposition::operator()(int j) const {
    return weights[j];
}

int clebsch::decomposition::multiplicity(const weight &irrep) const {
    assert(irrep.N == N);
    std::vector<clebsch::weight>::const_iterator it
        = std::lower_bound(weights.begin(), weights.end(), irrep);

    return it != weights.end() && *it == irrep ? multiplicities[it - weights.begin()] : 0;
}

// implementation of "index_adapter" starts here

clebsch::index_adapter::index_adapter(const clebsch::decomposition &decomp) :
        N(decomp.N),
        factor1(decomp.factor1.index()),
        factor2(decomp.factor2.index()) {
    for (int i = 0, s = decomp.size(); i < s; ++i) {
        indices.push_back(decomp(i).index());
        multiplicities.push_back(decomp.multiplicity(decomp(i)));
    }
}

int clebsch::index_adapter::size() const {
    return indices.size();
}

int clebsch::index_adapter::operator()(int j) const {
    return indices[j];
}

int clebsch::index_adapter::multiplicity(int irrep) const {
    std::vector<int>::const_iterator it = std::lower_bound(indices.begin(), indices.end(), irrep);

    return it != indices.end() && *it == irrep ? multiplicities[it - indices.begin()] : 0;
}

// implementation of "clebsch" starts here

void clebsch::coefficients::set(int factor1_state,
                                int factor2_state,
                                int multiplicity_index,
                                int irrep_state,
                                double value) {
    assert(0 <= factor1_state && factor1_state < factor1_dimension);
    assert(0 <= factor2_state && factor2_state < factor2_dimension);
    assert(0 <= multiplicity_index && multiplicity_index < multiplicity);
    assert(0 <= irrep_state && irrep_state < irrep_dimension);

    int coefficient_label[] = { factor1_state,
                                factor2_state,
                                multiplicity_index,
                                irrep_state };
    clzx[std::vector<int>(coefficient_label, coefficient_label
            + sizeof coefficient_label / sizeof coefficient_label[0])] = value;
}

void clebsch::coefficients::highest_weight_normal_form() {
    int hws = irrep_dimension - 1;

    // bring CGCs into reduced row echelon form
    for (int h = 0, i = 0; h < multiplicity - 1 && i < factor1_dimension; ++i) {
        for (int j = 0; h < multiplicity - 1 && j < factor2_dimension; ++j) {
            int k0 = h;

            for (int k = h + 1; k < multiplicity; ++k) {
                if (fabs((*this)(i, j, k, hws)) > fabs((*this)(i, j, k0, hws))) {
                    k0 = k;
                }
            }

            if ((*this)(i, j, k0, hws) < -EPS) {
                for (int i2 = i; i2 < factor1_dimension; ++i2) {
                    for (int j2 = i2 == i ? j : 0; j2 < factor2_dimension; ++j2) {
                        set(i2, j2, k0, hws, -(*this)(i2, j2, k0, hws));
                    }
                }
            } else if ((*this)(i, j, k0, hws) < EPS) {
                continue;
            }

            if (k0 != h) {
                for (int i2 = i; i2 < factor1_dimension; ++i2) {
                    for (int j2 = i2 == i ? j : 0; j2 < factor2_dimension; ++j2) {
                        double x = (*this)(i2, j2, k0, hws);
                        set(i2, j2, k0, hws, (*this)(i2, j2, h, hws));
                        set(i2, j2, h, hws, x);
                    }
                }
            }

            for (int k = h + 1; k < multiplicity; ++k) {
                for (int i2 = i; i2 < factor1_dimension; ++i2) {
                    for (int j2 = i2 == i ? j : 0; j2 < factor2_dimension; ++j2) {
                        set(i2, j2, k, hws, (*this)(i2, j2, k, hws) - (*this)(i2, j2, h, hws)
                                * (*this)(i, j, k, hws) / (*this)(i, j, h, hws));
                    }
                }
            }

            // next 3 lines not strictly necessary, might improve numerical stability
            for (int k = h + 1; k < multiplicity; ++k) {
                set(i, j, k, hws, 0.0);
            }

            ++h;
        }
    }

    // Gram-Schmidt orthonormalization
    for (int h = 0; h < multiplicity; ++h) {
        for (int k = 0; k < h; ++k) {
            double overlap = 0.0;
            for (int i = 0; i < factor1_dimension; ++i) {
                for (int j = 0; j < factor2_dimension; ++j) {
                    overlap += (*this)(i, j, h, hws) * (*this)(i, j, k, hws);
                }
            }

            for (int i = 0; i < factor1_dimension; ++i) {
                for (int j = 0; j < factor2_dimension; ++j) {
                    set(i, j, h, hws, (*this)(i, j, h, hws) - overlap * (*this)(i, j, k, hws));
                }
            }
        }

        double norm = 0.0;
        for (int i = 0; i < factor1_dimension; ++i) {
            for (int j = 0; j < factor2_dimension; ++j) {
                norm += (*this)(i, j, h, hws) * (*this)(i, j, h, hws);
            }
        }
        norm = std::sqrt(norm);

        for (int i = 0; i < factor1_dimension; ++i) {
            for (int j = 0; j < factor2_dimension; ++j) {
                set(i, j, h, hws, (*this)(i, j, h, hws) / norm);
            }
        }
    }
}

void clebsch::coefficients::compute_highest_weight_coeffs() {
    if (multiplicity == 0) {
        return;
    }

    std::vector<std::vector<int> > map_coeff(factor1_dimension,
                                             std::vector<int>(factor2_dimension, -1));
    std::vector<std::vector<int> > map_states(factor1_dimension,
                                              std::vector<int>(factor2_dimension, -1));
    int n_coeff = 0, n_states = 0;
    pattern p(factor1, 0);

    for (int i = 0; i < factor1_dimension; ++i, ++p) {
        weight pw(p.get_weight());
        pattern q(factor2, 0);
        for (int j = 0; j < factor2_dimension; ++j, ++q) {
            if (pw + q.get_weight() == irrep) {
                map_coeff[i][j] = n_coeff++;
            }
        }
    }

    if (n_coeff == 1) {
        for (int i = 0; i < factor1_dimension; ++i) {
            for (int j = 0; j < factor2_dimension; ++j) {
                if (map_coeff[i][j] >= 0) {
                    set(i, j, 0, irrep_dimension - 1, 1.0);
                    return;
                }
            }
        }
    }

    double *hw_system = new double[n_coeff * (factor1_dimension * factor2_dimension)];
    assert(hw_system != NULL);
    memset(hw_system, 0, n_coeff * (factor1_dimension * factor2_dimension) * sizeof (double));

    pattern r(factor1, 0);
    for (int i = 0; i < factor1_dimension; ++i, ++r) {
        pattern q(factor2, 0);

        for (int j = 0; j < factor2_dimension; ++j, ++q) {
            if (map_coeff[i][j] >= 0) {
                for (int l = 1; l <= N - 1; ++l) {
                    for (int k = 1; k <= l; ++k) {
                        if ((k == 1 || r(k, l) + 1 <= r(k - 1, l - 1)) && r(k, l) + 1 <= r(k, l + 1)) {
                            ++r(k, l);
                            int h = r.index();
                            --r(k, l);

                            if (map_states[h][j] < 0) {
                                map_states[h][j] = n_states++;
                            }

                            hw_system[n_coeff * map_states[h][j] + map_coeff[i][j]]
                                += r.raising_coeff(k, l);
                        }

                        if ((k == 1 || q(k, l) + 1 <= q(k - 1, l - 1)) && q(k, l) + 1 <= q(k, l + 1)) {
                            ++q(k, l);
                            int h = q.index();
                            --q(k, l);

                            if (map_states[i][h] < 0) {
                                map_states[i][h] = n_states++;
                            }


                            hw_system[n_coeff * map_states[i][h] + map_coeff[i][j]]
                                += q.raising_coeff(k, l);
                        }
                    }
                }
            }
        }
    }

    int lwork = -1, info;
    double worksize;

    double *singval = new double[std::min(n_coeff, n_states)];
    assert(singval != NULL);
    double *singvec = new double[n_coeff * n_coeff];
    assert(singvec != NULL);

    dgesvd_("A",
            "N",
            &n_coeff,
            &n_states,
            hw_system,
            &n_coeff,
            singval,
            singvec,
            &n_coeff,
            NULL,
            &n_states,
            &worksize,
            &lwork,
            &info);
    assert(info == 0);

    lwork = worksize;
    double *work = new double[lwork];
    assert(work != NULL);

    dgesvd_("A",
            "N",
            &n_coeff,
            &n_states,
            hw_system,
            &n_coeff,
            singval,
            singvec,
            &n_coeff,
            NULL,
            &n_states,
            work,
            &lwork,
            &info);
    assert(info == 0);

    for (int i = 0; i < multiplicity; ++i) {
        for (int j = 0; j < factor1_dimension; ++j) {
            for (int k = 0; k < factor2_dimension; ++k) {
                if (map_coeff[j][k] >= 0) {
                    double x = singvec[n_coeff * (n_coeff - 1 - i) + map_coeff[j][k]];

                    if (fabs(x) > EPS) {
                        set(j, k, i, irrep_dimension - 1, x);
                    }
                }
            }
        }
    }

    // uncomment next line to bring highest-weight coefficients into "normal form"
    // highest_weight_normal_form();

    delete[] work;
    delete[] singvec;
    delete[] singval;
    delete[] hw_system;
}

void clebsch::coefficients::compute_lower_weight_coeffs(int multip_index,
                                                        int state,
                                                        std::vector<char> &done) {
    weight statew(pattern(irrep, state).get_weight());
    pattern p(irrep, 0);
    std::vector<int> map_parent(irrep_dimension, -1),
                     map_multi(irrep_dimension, -1),
                     which_l(irrep_dimension, -1);
    int n_parent = 0, n_multi = 0;

    for (int i = 0; i < irrep_dimension; ++i, ++p) {
        weight v(p.get_weight());

        if (v == statew) {
            map_multi[i] = n_multi++;
        } else for (int l = 1; l < N; ++l) {
            --v(l);
            ++v(l + 1);
            if (v == statew) {
                map_parent[i] = n_parent++;
                which_l[i] = l;
                if (!done[i]) {
                    compute_lower_weight_coeffs(multip_index, i, done);
                }
                break;
            }
            --v(l + 1);
            ++v(l);
        }
    }

    double *irrep_coeffs = new double[n_parent * n_multi];
    assert(irrep_coeffs != NULL);
    memset(irrep_coeffs, 0, n_parent * n_multi * sizeof (double));

    double *prod_coeffs = new double[n_parent * factor1_dimension * factor2_dimension];
    assert(prod_coeffs != NULL);
    memset(prod_coeffs, 0, n_parent * factor1_dimension * factor2_dimension * sizeof (double));

    std::vector<std::vector<int> > map_prodstat(factor1_dimension,
                                                std::vector<int>(factor2_dimension, -1));
    int n_prodstat = 0;

    pattern r(irrep, 0);
    for (int i = 0; i < irrep_dimension; ++i, ++r) {
        if (map_parent[i] >= 0) {
            for (int k = 1, l = which_l[i]; k <= l; ++k) {
                if (r(k, l) > r(k + 1, l + 1) && (k == l || r(k, l) > r(k, l - 1))) {
                    --r(k, l);
                    int h = r.index();
                    ++r(k, l);

                    irrep_coeffs[n_parent * map_multi[h] + map_parent[i]] += r.lowering_coeff(k, l);
                }
            }

            pattern q1(factor1, 0);
            for (int j1 = 0; j1 < factor1_dimension; ++j1, ++q1) {
                pattern q2(factor2, 0);

                for (int j2 = 0; j2 < factor2_dimension; ++j2, ++q2) {
                    if (std::fabs((*this)(j1, j2, multip_index, i)) > EPS) {
                        for (int k = 1, l = which_l[i]; k <= l; ++k) {
                            if (q1(k, l) > q1(k + 1, l + 1) && (k == l || q1(k, l) > q1(k, l - 1))) {
                                --q1(k, l);
                                int h = q1.index();
                                ++q1(k, l);

                                if (map_prodstat[h][j2] < 0) {
                                    map_prodstat[h][j2] = n_prodstat++;
                                }

                                prod_coeffs[n_parent * map_prodstat[h][j2] + map_parent[i]] +=
                                        (*this)(j1, j2, multip_index, i) * q1.lowering_coeff(k, l);
                            }

                            if (q2(k, l) > q2(k + 1, l + 1) && (k == l || q2(k, l) > q2(k, l - 1))) {
                                --q2(k, l);
                                int h = q2.index();
                                ++q2(k, l);

                                if (map_prodstat[j1][h] < 0) {
                                    map_prodstat[j1][h] = n_prodstat++;
                                }

                                prod_coeffs[n_parent * map_prodstat[j1][h] + map_parent[i]] +=
                                        (*this)(j1, j2, multip_index, i) * q2.lowering_coeff(k, l);
                            }
                        }
                    }
                }
            }
        }
    }

    double worksize;
    int lwork = -1, info;

    dgels_("N",
           &n_parent,
           &n_multi,
           &n_prodstat,
           irrep_coeffs,
           &n_parent,
           prod_coeffs,
           &n_parent,
           &worksize,
           &lwork,
           &info);
    assert(info == 0);

    lwork = worksize;
    double *work = new double[lwork];
    assert(work != NULL);

    dgels_("N",
           &n_parent,
           &n_multi,
           &n_prodstat,
           irrep_coeffs,
           &n_parent,
           prod_coeffs,
           &n_parent,
           work,
           &lwork,
           &info);
    assert(info == 0);

    for (int i = 0; i < irrep_dimension; ++i) {
        if (map_multi[i] >= 0) {
            for (int j = 0; j < factor1_dimension; ++j) {
                for (int k = 0; k < factor2_dimension; ++k) {
                    if (map_prodstat[j][k] >= 0) {
                        double x = prod_coeffs[n_parent * map_prodstat[j][k] + map_multi[i]];

                        if (fabs(x) > EPS) {
                            set(j, k, multip_index, i, x);
                        }
                    }
                }
            }

            done[i] = true;
        }
    }

    delete[] work;
    delete[] prod_coeffs;
    delete[] irrep_coeffs;
}

clebsch::coefficients::coefficients(const weight &irrep, const weight &factor1, const weight &factor2) :
        N(irrep.N),
        factor1(factor1),
        factor2(factor2),
        irrep(irrep),
        factor1_dimension(factor1.dimension()),
        factor2_dimension(factor2.dimension()),
        irrep_dimension(irrep.dimension()),
        multiplicity(decomposition(factor1, factor2).multiplicity(irrep)) {
    assert(factor1.N == irrep.N);
    assert(factor2.N == irrep.N);

    compute_highest_weight_coeffs();

    for (int i = 0; i < multiplicity; ++i) {
        std::vector<char> done(irrep_dimension, 0);
        done[irrep_dimension - 1] = true;
        for (int j = irrep_dimension - 1; j >= 0; --j) {
            if (!done[j]) {
                compute_lower_weight_coeffs(i, j, done);
            }
        }
    }
}

double clebsch::coefficients::operator()(int factor1_state,
                                         int factor2_state,
                                         int multiplicity_index,
                                         int irrep_state) const {
    assert(0 <= factor1_state && factor1_state < factor1_dimension);
    assert(0 <= factor2_state && factor2_state < factor2_dimension);
    assert(0 <= multiplicity_index && multiplicity_index < multiplicity);
    assert(0 <= irrep_state && irrep_state < irrep_dimension);

    int coefficient_label[] = { factor1_state,
                                factor2_state,
                                multiplicity_index,
                                irrep_state };
    std::map<std::vector<int>, double>::const_iterator it(
            clzx.find(std::vector<int>(coefficient_label, coefficient_label
                    + sizeof coefficient_label / sizeof coefficient_label[0])));

    return it != clzx.end() ? it->second : 0.0;
}

// sample driver routine

using namespace std;

int main() {
    while (true) {
        int choice, N;

        cout << "What would you like to do?" << endl;
        cout << "1) Translate an i-weight S to its index P(S)" << endl;
        cout << "2) Recover an i-weight S from its index P(S)" << endl;
        cout << "3) Translate a pattern M to its index Q(M)" << endl;
        cout << "4) Recover a pattern M from its index Q(M)" << endl;
        cout << "5) Calculate Clebsch-Gordan coefficients for S x S' -> S''" << endl;
        cout << "6) Calculate all Glebsch-Gordan coefficients for S x S'" << endl;
        cout << "0) Quit" << endl;

        do {
            cin >> choice;
        } while (choice < 0 || choice > 6);

        if (choice == 0) {
            break;
        }

        cout << "N (e.g. 3): ";
        cin >> N;

        switch (choice) {
            case 1: {
                clebsch::weight S(N);
                cout << "Irrep S: ";
                for (int k = 1; k <= N; ++k) {
                    cin >> S(k);
                }
                cout << S.index() << endl;
                break;
            }
            case 2: {
                int P;
                cout << "Index: ";
                cin >> P;
                clebsch::weight S(N, P);
                cout << "I-weight:";
                for (int k = 1; k <= N; ++k) {
                    cout << ' ' << S(k);
                }
                cout << endl;
                break;
            }
            case 3: {
                clebsch::pattern M(N);
                for (int l = N; l >= 1; --l) {
                    cout << "Row l = " << l << ": ";
                    for (int k = 1; k <= l; ++k) {
                        cin >> M(k, l);
                    }
                }
                cout << "Index: " << M.index() + 1 << endl;
                break;
            }
            case 4: {
                clebsch::weight S(N);
                cout << "Irrep S: ";
                for (int i = 1; i <= N; ++i) {
                    cin >> S(i);
                }

                int Q;
                cout << "Index (1..dim(S)): ";
                cin >> Q;
                clebsch::pattern M(S, Q - 1);
                for (int l = N; l >= 1; --l) {
                    for (int k = 1; k <= l; ++k) {
                        cout << M(k, l) << '\t';
                    }
                    cout << endl;
                }
                break;
            }
            case 5: {
                clebsch::weight S(N);
                cout << "Irrep S (e.g.";
                for (int k = N - 1; k >= 0; --k) {
                    cout << ' ' << k;
                }
                cout << "): ";
                for (int k = 1; k <= N; ++k) {
                    cin >> S(k);
                }

                clebsch::weight Sprime(N);
                cout << "Irrep S' (e.g.";
                for (int k = N - 1; k >= 0; --k) {
                    cout << ' ' << k;
                }
                cout << "): ";
                for (int k = 1; k <= N; ++k) {
                    cin >> Sprime(k);
                }

                clebsch::decomposition decomp(S, Sprime);
                cout << "Littlewood-Richardson decomposition S \\otimes S' = \\oplus S'':" << endl;
                cout << "[irrep index] S'' (outer multiplicity) {dimension d_S}" << endl;
                for (int i = 0; i < decomp.size(); ++i) {
                    cout << "[" << decomp(i).index() << "] ";
                    for (int k = 1; k <= N; ++k) {
                        cout << decomp(i)(k) << ' ';
                    }
                    cout << '(' << decomp.multiplicity(decomp(i)) << ") {"
                         << decomp(i).dimension() << "}" << endl;;
                }

                clebsch::weight Sdoubleprime(N);
                for (bool b = true; b; ) {
                    cout << "Irrep S'': ";
                    for (int k = 1; k <= N; ++k) {
                        cin >> Sdoubleprime(k);
                    }
                    for (int i = 0; i < decomp.size(); ++i) {
                        if (decomp(i) == Sdoubleprime) {
                            b = false;
                            break;
                        }
                    }
                    if (b) {
                        cout << "S'' does not occur in the decomposition" << endl;
                    }
                }

                int alpha;
                while (true) {
                    cout << "Outer multiplicity index: ";
                    cin >> alpha;
                    if (1 <= alpha && alpha <= decomp.multiplicity(Sdoubleprime)) {
                        break;
                    }
                    cout << "S'' does not have this multiplicity" << endl;
                }

                string file_name;
                cout << "Enter file name to write to file (leave blank for screen output): ";
                cin.ignore(1234, '\n');
                getline(cin, file_name);

                const clebsch::coefficients C(Sdoubleprime, S, Sprime);
                int dimS = S.dimension(),
                    dimSprime = Sprime.dimension(),
                    dimSdoubleprime = Sdoubleprime.dimension();

                ofstream os(file_name.c_str());
                (file_name.empty() ? cout : os).setf(ios::fixed);
                (file_name.empty() ? cout : os).precision(15);
                (file_name.empty() ? cout : os) << "List of nonzero CGCs for S x S' => S'', alpha" << endl;
                (file_name.empty() ? cout : os) << "Q(M)\tQ(M')\tQ(M'')\tCGC" << endl;
                for (int i = 0; i < dimSdoubleprime; ++i) {
                    for (int j = 0; j < dimS; ++j) {
                        for (int k = 0; k < dimSprime; ++k) {
                            double x = double(C(j, k, alpha - 1, i));

                            if (fabs(x) > clebsch::EPS) {
                                (file_name.empty() ? cout : os) << j + 1 << '\t'
                                    << k + 1 << '\t' << i + 1 << '\t' << x << endl;
                            }
                        }
                    }
                }

                break;
            }
            case 6: {
                clebsch::weight S(N);
                cout << "Irrep S (e.g.";
                for (int k = N - 1; k >= 0; --k) {
                    cout << ' ' << k;
                }
                cout << "): ";
                for (int k = 1; k <= N; ++k) {
                    cin >> S(k);
                }

                clebsch::weight Sprime(N);
                cout << "Irrep S' (e.g.";
                for (int k = N - 1; k >= 0; --k) {
                    cout << ' ' << k;
                }
                cout << "): ";
                for (int k = 1; k <= N; ++k) {
                    cin >> Sprime(k);
                }

                string file_name;
                cout << "Enter file name to write to file (leave blank for screen output): ";
                cin.ignore(1234, '\n');
                getline(cin, file_name);

                ofstream os(file_name.c_str());
                (file_name.empty() ? cout : os).setf(ios::fixed);
                (file_name.empty() ? cout : os).precision(15);

                clebsch::decomposition decomp(S, Sprime);
                (file_name.empty() ? cout : os) <<
                    "Littlewood-Richardson decomposition S \\otimes S' = \\oplus S'':" << endl;
                (file_name.empty() ? cout : os) <<
                    "[irrep index] S'' (outer multiplicity) {dimension d_S}" << endl;
                for (int i = 0; i < decomp.size(); ++i) {
                    (file_name.empty() ? cout : os) << "[" << decomp(i).index() << "] ";
                    for (int k = 1; k <= N; ++k) {
                        (file_name.empty() ? cout : os) << decomp(i)(k) << ' ';
                    }
                    (file_name.empty() ? cout : os) << '(' << decomp.multiplicity(decomp(i)) << ") {"
                         << decomp(i).dimension() << "}" << endl;;
                }

                for (int i = 0; i < decomp.size(); ++i) {
                    const clebsch::coefficients C(decomp(i),S, Sprime);
                    int dimS = S.dimension(),
                        dimSprime = Sprime.dimension(),
                        dimSdoubleprime = decomp(i).dimension();

                    for (int m = 0; m < C.multiplicity; ++m) {
                        (file_name.empty() ? cout : os) << "List of nonzero CGCs for S x S' => S'' = (";
                        for (int j = 1; j <= N; ++j) cout << decomp(i)(j) << (j < N ? ' ' : ')');
                        (file_name.empty() ? cout : os) << ", alpha = " << m + 1 << endl;
                        (file_name.empty() ? cout : os) << "Q(M)\tQ(M')\tQ(M'')\tCGC" << endl;
                        for (int i = 0; i < dimSdoubleprime; ++i) {
                            for (int j = 0; j < dimS; ++j) {
                                for (int k = 0; k < dimSprime; ++k) {
                                    double x = double(C(j, k, m, i));

                                    if (fabs(x) > clebsch::EPS) {
                                        (file_name.empty() ? cout : os) << j  + 1<< '\t'
                                            << k + 1 << '\t' << i + 1 << '\t' << x << endl;
                                    }
                                }
                            }
                        }

                        (file_name.empty() ? cout : os) << endl;
                    }
                }

                break;
            }
        }
    }

    return 0;
}
\end{lstlisting}
\end{document}